\documentclass[useAMS,usenatbib]{mn2e}
\usepackage{graphicx}
\usepackage{amssymb}
\title[Short GRB X-ray flares]{X-ray flare candidates in short gamma-ray bursts}
\author[R. Margutti et al.]{R. Margutti$^{1,7}$ \thanks{E-mail:
raffaella.margutti@brera.inaf.it (RM)}, G. Chincarini$^{1,2}$, J. {Granot}$^{3,4,5}$, C. {Guidorzi}$^{6}$, E. {Berger}$^{7}$ \and
M.G. {Bernardini}$^{1}$, N. {Gehrels}$^{8}$, A.M. {Soderberg}$^{7}$, M. {Stamatikos}$^{8}$, E. {Zaninoni}$^{1,10}$\\
$^{1}$ INAF Osservatorio Astronomico di Brera, via Bianchi 46, Merate 23807, Italy \\
$^{2}$ Univerisit\'a Milano Bicocca, Dip. Fisica G. Occhialini, P.zza della Scienza 3, Milano 20126, Italy\\ 
$^{3}$ Racah Institute of Physics, The Hebrew University, Jerusalem 91904, Israel \\
$^{4}$ Raymond and Beverly Sackler School of Physics \& Astronomy, Tel Aviv University, Tel Aviv 69978, Israel\\
$^{5}$ Centre for Astrophysics Research, University of Hertfordshire, College Lane, HatÞeld, Herts, AL109AB, UK \\
$^{6}$ University of Ferrara, Physics Dept., via Saragat 1, I-44122 Ferarra, Italy \\
$^{7}$ Harvard-Smithsonian Center for Astrophysics, 60 Garden Street, Cambridge, MA02138\\
$^{8}$ NASA-Goddard Space Flight Center, Greenbelt, Maryland 20771\\
$^{9}$ Dept. of Physics and Center for Cosmology and Astro-Particle Physics, Ohio State University, Columbus, OH 43210, USA\\
$^{10}$ Univerisit\'a di Padova, Dip. Astronomia, v. dell' Osservatorio 31, Padova 35122, Italy\\}
\begin{document}

\date{Accepted Year Month Day. Received Year Month Day; in original form Year Month Day}
\pagerange{\pageref{firstpage}--\pageref{lastpage}} \pubyear{2010}
\maketitle
\label{firstpage}

\begin{abstract}
	We present the first systematic study of X-ray flare candidates in short gamma-ray bursts (SGRBs) 
	exploiting the large 6-year \emph{Swift} database with the aim to constrain the physical nature of such fluctuations.
	We find that flare candidates appear in different types of SGRB host galaxy environments
	and show no clear correlation with the X-ray afterglow lifetime; flare candidates are detected both in
	SGRBs with a bright extended emission in the soft $\gamma$-rays and in SGRBs
	which do not show such component.
	We furthermore show that SGRB X-ray flare candidates only \emph{partially} 
	share the set of observational properties of long GRB (LGRB) flares. In particular,
	the main parameter driving the duration evolution of X-ray variability episodes in both classes
	is found to be the elapsed time from the explosion, with very limited dependence on the different
	progenitors, environments, central engine life-times, prompt variability time-scales and energy budgets.
	On the contrary, SGRB flare candidates significantly differ from LGRB flares in terms of 
	peak luminosity, isotropic energy, flare-to-prompt luminosity ratio and relative variability flux. 
	However, these differences disappear when the central engine time-scales and energy budget are
	accounted for, suggesting that (i) flare candidates and prompt pulses in SGRBs likely have a common origin;
	(ii) similar dissipation and/or emission mechanisms are responsible for the prompt and flare emission in long and
	short GRBs, with SGRBs being less energetic albeit faster evolving versions of the long class.  
	Finally, we show that in strict analogy to the SGRB prompt
	emission, flares candidates fall off the lag-luminosity relation defined by LGRBs, thus strengthening 
	the SGRB flare-prompt pulse connection.
\end{abstract}

\section{Introduction}
\label{Sec:Int}
With an isotropic peak luminosity up to $10^{54}\,\rm{erg\,s^{-1}}$, 
gamma-ray bursts (GRBs) are the brightest objects in the $\gamma$-ray sky during their short lives ($\Delta t \sim 0.1-100$ s).
Their duration-spectral hardness distribution gives evidence for the presence of two 
classes (\citealt{Kouveliotou93}): long  and short GRBs (observed duration longer and shorter than 2 s,
respectively), with short 
bursts appearing slightly harder. The dichotomy in the duration-hardness dimensions suggested separate
progenitor populations. However, until a few years ago, the distances, energy and environments of 
SGRBs (short GRBs) remained highly uncertain due to the poor localisation. 

The breakthrough in the study of SGRBs occurred thanks to the rapid slew capabilities of the \emph{Swift}
spacecraft (\citealt{Gehrels04}) which allowed spectroscopic observations to be performed at very early times.
These observations revealed that SGRBs are cosmological, 
with prompt luminosities comparable to LGRBs albeit significantly less energetic; with similar afterglows 
(\citealt{Nysewander09}) but residing in completely different environments.
In sharp contrast to LGRBs, short bursts have been localised both in early-type and late-type host 
galaxies (see \citealt{Berger11} and references therein), pointing to an \emph{old} progenitor
population. The detection of supernovae associated to LGRBs (see e.g. \citealt{Kulkarni98};
\citealt{Stanek03}; \citealt{Fruchter06} and references therein) provided instead support to models invoking \emph{young} stellar 
progenitors. According to the standard scenario, LGRBs originate from the collapse of rapidly-rotating, 
massive stars (\citealt{MacFadyen99}), while  SGRBs are believed to result from the coalescence of
a binary system of compact objects (neutron star plus neutron star NS+NS or neutron star plus black hole
NS+BH, \citealt{Paczynski86}; \citealt{Eichler89}; \citealt{Narayan92}).

Despite fundamental theoretical and observational progress, the nature of SGRB progenitors 
remains elusive.  Numerical simulations show that the active stage of a NS+NS merger typically
lasts $\sim(0.01-0.1)$ s  
(see e.g. \citealt{Nakar07} and references therein\footnote{In a recent study \cite{Rezzolla11} found $\Delta t\sim 0.3$ s.}): 
material ejected during the merger is expected to accrete
on time-scales of the order of $1-10$ s (the exact value depending on the accreting disk viscosity 
parameter and details of the ejection process). Thus, the detection of central engine activity on time-scales much longer than the
usual dynamical or even viscous time-scales  would 
challenge the currently accepted scenario (see \citealt{Nakar07} for a recent review).

Long-lasting ($\Delta t \gg 10\,\rm{s}$), soft energy tails detected in several SGRBs during their 
prompt emission  (the so-called extended emission, see \citealt{Norris10a} and references therein) 
represent such a case and pose severe constraints to existing models, especially when energetically dominating 
with respect to the primary burst (\citealt{Perley09}).  The same is true for the recently discovered presence
of precursors (\citealt{Troja10}).
Equally challenging would be the detection 
of late-time central engine activity in the form of \emph{flares} superimposed over the smooth SGRB X-ray afterglow.

Flares are currently detected in  $\sim30$\% of \emph{long} GRBs X-ray afterglows (\citealt{Chincarini10}) as 
fast-rise exponential-decay features whose spectral and temporal properties have been demonstrated to
show a strict analogy to LGRB prompt pulses (\citealt{Margutti10b}): this finding suggested that flares might
originate from re-activations of the LGRB central engine. Several ideas on how to explain the possible 
presence of flares in \emph{short} GRBs have been explored as well: 
the fragmentation of the outer parts of an hyper-accreting disk around the newly formed 
black hole as a result of gravitational instabilities could potentially lead to large-amplitude 
variations of the central engine output of both long and short GRBs (\citealt{Perna06}). Alternatively,
the late-time accretion of material launched into eccentric but gravitationally bound orbits
during the compact binary merger could provide the fuel to revive the central engine activity 
(\citealt{Rosswog07}). The long term evolution of debris following the tidal disruption of
compact objects has been identified by \cite{Lee09} as a feasible mechanism to produce flares.
Finally, as an alternative  in the context of accretion-powered models, magnetic
halting may also give rise to secondary episodes of delayed activity as suggested by \cite{Proga06}.
However, the observational properties of flares in SGRBs have not been determined, yet, so that
it is at the moment unclear if any of these models would be able to explain the observations.

While SGRB X-ray light-curves clearly show temporal variability superimposed over a smooth 
decay, the presence of real flares in short bursts is questionable. In particular, it is at the moment
unclear if what is currently identified as \emph{short} GRB flare emission (see e.g. \citealt{LaParola06}
for GRB\,051210) quantitatively shares the very same properties of the population of  
\emph{long} GRB flares: are there fast varying $\Delta t/t\ll1$, prominent temporal features in the
afterglow of SGRBs with properties reminiscent of the long GRB flaring emission? 
Do SGRB flare \emph{candidates} follow the entire set of relations 
found from the analysis of real flares in long bursts? In particular: is the evolution of their temporal
and energetic properties compatible with the flare-like behaviour identified by \cite{Chincarini10}?
What is the typical amount of energy released during such episodes of variability?
Is there any link between the late-time variability which appears in the X-ray afterglow of SGRBs
and their prompt emission? Negligible spectral lag is a defining characteristic of 
SGRB prompt pulses: is this picture still valid when considering their late-time variability? 

Prompted by this set of still open questions, we present the first systematic study of X-ray flare 
candidates in \emph{short} GRBs, taking advantage from the large \emph{Swift} 6-year data-base. 
Through a homogeneous temporal and spectral analysis of the widest sample of SGRB 
light-curves available at the time of writing, this study allows us to perform a one-to-one comparison 
with the properties of X-ray flares detected in long bursts (\citealt{Chincarini10}, \citealt{Margutti10b}, 
\citealt{Bernardini11}, \citealt{Margutti11}). The primary goal of this paper is to observationally 
constrain the origin of SGRB flare candidates providing the reader with a complete picture 
of their properties. 

This work is organised as follows: the sample selection and data reduction is presented in
Section \ref{Sec:DataAn}. Results are described in Section \ref{Sec:Res} and discussed in Section \ref{Sec:Disc}.
Conclusions are drawn in Section  \ref{Sec:Con}. 

The GRB phenomenology is presented 
in the observer frame.  Isotropic equivalent luminosities and energies
are listed. The observer frame 0.3-10 keV energy band is adopted unless specified.
The zero time is assumed to be the trigger time. We use the notation: $Y^{\rm{SGRB}}_{\rm{F}}$ 
($Y^{\rm{LGRB}}_{\rm{P}}$) to indicate that $Y$ refers to the flare (prompt) emission of SGRBs (LGRBs).
All the quoted uncertainties are given at 68\% confidence
level (c.l.). Standard cosmological quantities have been adopted: 
$H_{0}=70\,\rm{Km\,s^{-1}\,Mpc^{-1}}$, $\Omega_{\Lambda }=0.7$, $\Omega_{\rm{M}}=0.3$.

\section{sample selection and data analysis }
\label{Sec:DataAn}
We select the short GRBs detected by the \emph{Swift} Burst Alert Telescope (BAT, \citealt{Barthelmy05a}) and promptly 
re-pointed by the \emph{Swift} X-Ray Telescope (XRT, \citealt{Burrows05})  between April 2005 and February 2011.
The short nature of each event is established using the combined information from the duration, hardness and spectral
lag of its prompt $\gamma$-ray emission: a  prompt $\gamma$-ray duration $T_{90}\lesssim 2$ s coupled to a hard $\gamma$-ray 
emission with photon index $\Gamma \lesssim 1.5$ and a negligible $\gamma$-ray spectral lag $\tau^{\gamma}_{\rm{lag}}$  are 
considered indicative of a \emph{short} GRB nature (see Table \ref{Tab:sample}). The morphology of the host galaxy 
is also used as an additional indicator, when available. The final sample comprises 60 SGRBs.
The presence of X-ray variability in each SGRB is investigated following 
the method by \cite{Margutti11}, used to determine the presence of flares in long GRBs. Only GRBs showing
fluctuations with a minimum $2\,\sigma$\footnote{A $3\sigma$ threshold would only
exclude GRB\,051210, where the fluctuation has a significance of $\sim2.8\sigma$.} significance with respect to the continuum have been considered  in the
following analysis. This procedure automatically identifies the best time intervals to be searched for 
the presence of X-ray flare candidates in SGRBs. Out of $\sim$60 \emph{Swift} SGRBs, 8 satisfy the variability requirement 
above (Table \ref{Tab:sample})\footnote{The percentage of SGRBs with variable XRT light-curve $8/60\sim13\%$ is
much less than the $\sim30\%$ of LGRBs showing flares (\citealt{Chincarini10}). This result suggests that the percentage
of SGRB light-curves with variability superimposed is lower than in LGRBs. However, the lower statistics
characterising the SGRB curves prevents us from drawing firm conclusions. This topic will be addressed in a separate
work.}. 
Notably, the sample includes the unique 2 SGRBs with secure early-type host
identification: GRB\,050724 (\citealt{Barthelmy05b}) and GRB\,100117A  (\citealt{Fong2010}). In three cases
(GRB\,050724, GRB\,070724 and GRB\,071227, in boldface in Table \ref{Tab:sample} )
an extended emission (EE) has been detected in the soft gamma-ray
energy range after the short hard spike (\citealt{Norris10a}; \citealt{Norris11}). In the other cases, an upper limit on the 
EE to IPC (Initial Pulse Complex) intensity ratio ($R_{\rm{int}}\equiv EE_{\rm{int}}/IPC_{\rm{int}}$) has been provided by 
\cite{Norris10a}: for the sample of events without EE the upper limit on $R_{\rm{int}}$ is found to be
a factor $\gtrsim10$ below the typical $R_{\rm{int}}$ of SGRBs with detected EE (Table \ref{Tab:sample}, column 7).
Finally, GRB\,100816A has not been
included in the sample in spite of its $T_{90}=2.9 \pm 0.6$ s (\citealt{Markwardt10}) since the low statistics 
prevents the $\gamma$-ray lag analysis from giving definitive results on its 
possible short nature (\citealt{Norris10b}). The burst is however considered a SGRB in  \cite{Norris11}. 
\subsection{Swift-BAT data analysis}
\label{SubSec:BAT}
BAT data have been processed using standard Swift-BAT analysis tools within \textsc{heasoft} (v. 6.10).
In particular, the \textsc{batgrbproduct} script has been used to generate event lists and quality maps
necessary to construct $4\,\rm{ms}$ mask-weighted and background-subtracted light-curves
in the 50-100 keV and 100-200 keV anergy bands. The ground-refined coordinates provided by the
BAT-refined circulars have been adopted; standard filtering and screening criteria have been applied.
\subsection{Swift-XRT data analysis}
\label{SubSec:XRT}

\begin{figure}
\vskip -0.0 true cm
\centering
    \includegraphics[scale=0.5]{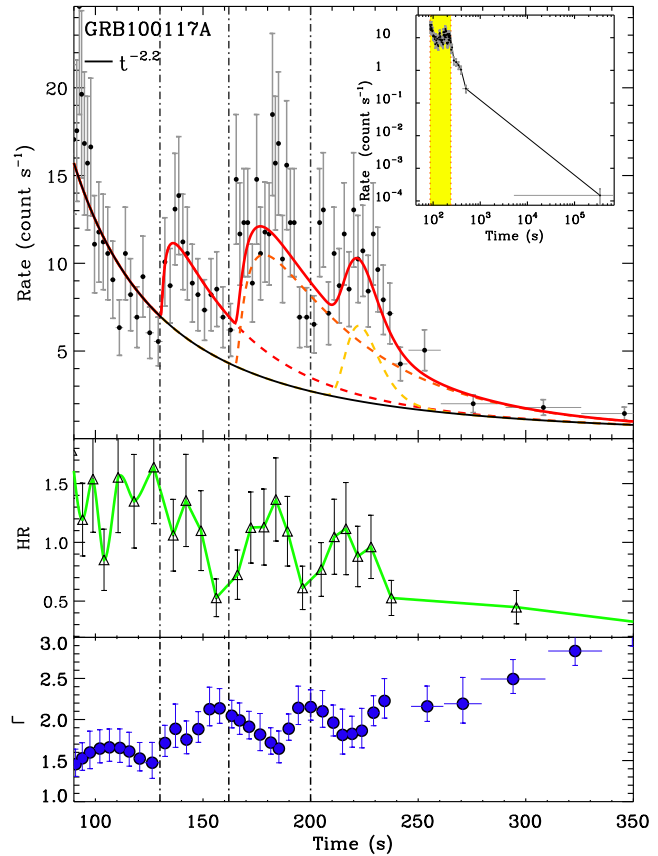}
     \caption{\emph{Upper panel:} 0.3-10 keV count-rate light-curve of GRB\,100117A. Black solid line:
       continuous  X-ray emission underlying the flare candidates computed as described in Section \ref{SubSec:XRT};
       dashed lines: best-fitting flare candidate emission; red solid line: best estimate of the total  
      emission. The vertical dot-dashed lines mark the flare candidate onset times.
	 \emph{Inset:} Complete \emph{Swift}-XRT light-curve. The yellow filled area marks the time window for the computation of the CCF lag
	 (Sect. \ref{SubSec:lagcomputation}).
       \emph{Middle panel:} hardness ratio (HR) evolution with time; the HR is computed between 1.5-10 keV (hard band) and 
       0.3-1.5 keV  (soft band). \emph{Lower panel}: Spectral photon index evolution with time as calculated by Evans et al., 2010.}
\label{Fig:lc100117A}
\end{figure}

\begin{figure}
\vskip -0.0 true cm
\centering
    \includegraphics[scale=0.38]{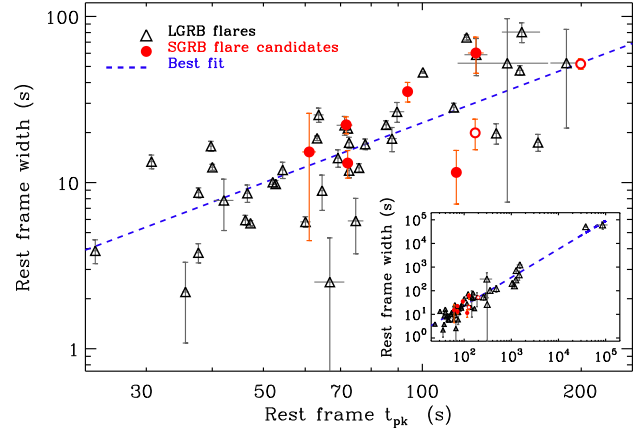}
      \caption{Rest frame width vs. peak time relation for LGRB early-time flares from Chincarini et al., 2010 
      (open triangles) and SGRB flare candidates with and without extended emission (red open and filled circles,
      respectively) . \emph{Inset:} Complete view of the $w/(1+z)$ vs. $t_{\rm{pk}}/(1+z)$ relation established by LGRB flares
     	obtained joining the data from Chincarini et al., 2010 and Bernardini et al., 2011. The blue dashed line in both
      plots	marks the best fitting relation calculated on LGRB flares: 
       $\frac{w}{1+z}=10^{(-1.0\pm0.5)}\big(\frac{t_{\rm{pk}}}{1+z}\big)^{(1.2\pm0.2)}$, where $w$ and $t_{\rm{pk}}$ are measured in
      seconds.}
\label{Fig:wtp}
\end{figure}

XRT data have been processed with the latest  \textsc{heasoft}  release available at the time of writing 
(v. 6.10) and corresponding calibration files: standard filtering and screening criteria have been applied.
Pile-up corrections have been applied when necessary (\citealt{Romano06}; \citealt{Vaughan06}). 
Count-rate light-curves have been extracted in the total XRT 0.3-10 keV energy band as well as in
the 0.3-1 keV, 3-10 keV, 0.3-1.5 keV, 1.5-10 keV and 4-10 keV energy bands. The 0.3-10 keV count-rate
light-curves have been re-binned at a minimum signal-to-noise ratio SN=4 and then searched for
statistically significant temporal variability superimposed over a smooth afterglow decay.  A two-step
procedure has been followed: first the smooth continuum contribution has been determined applying the method
by \cite{Margutti11}. A simple power-law or a smoothly joined broken power-law model is adopted
(black solid line of Fig. \ref{Fig:lc100117A}). As a second step, the properties of statistically significant
fluctuations with respect to the continuum have been determined adding a number of \cite{Norris05} profiles 
to the best fitting continuum model. The best fitting \cite{Norris05} profiles constitute the sample of
X-ray flare candidates of SGRBs analysed in this work. Figure \ref{Fig:lc100117A} shows GRB\,100117A
as an example: 3 distinct episodes of variability have been identified.
The best fitting parameters 
of the entire sample are listed in Table \ref{Tab:flarepar}. The choice of the  \cite{Norris05} profile allows us 
to perform a one-to-one  comparison with the properties of X-ray flares and prompt pulses in LGRBs 
(\citealt{Chincarini10}; \citealt{Bernardini11}): Fig. \ref{Fig:wtp} shows the evolution of the SGRB flare
candidates width with time compared to LGRB flares.

The evolution of the spectral properties of any source can be constrained through the analysis
of its hardness ratio (HR), which is here defined as 
$\rm{HR}=\frac{Counts(1.5-10\, \rm{keV})}{Counts(0.3-1.5\, \rm{keV})}$. A different binning with respect
to the total 0.3-10 keV light-curve has been used for the 1.5-10 keV and 0.3-1.5 keV light-curves to 
improve the HR signal-to-noise ratio. The temporal evolution of the spectral
photon index $\Gamma$ has been  calculated by \citealt{Evans10}. Results are portrayed in Fig. 
\ref{Fig:lc100117A}, \ref{Fig:lc050724051210}, \ref{Fig:lc051227060313}, \ref{Fig:lclast} and  \ref{Fig:lclastlast}.

\begin{figure}
\vskip -0.0 true cm
\centering
    \includegraphics[scale=0.4]{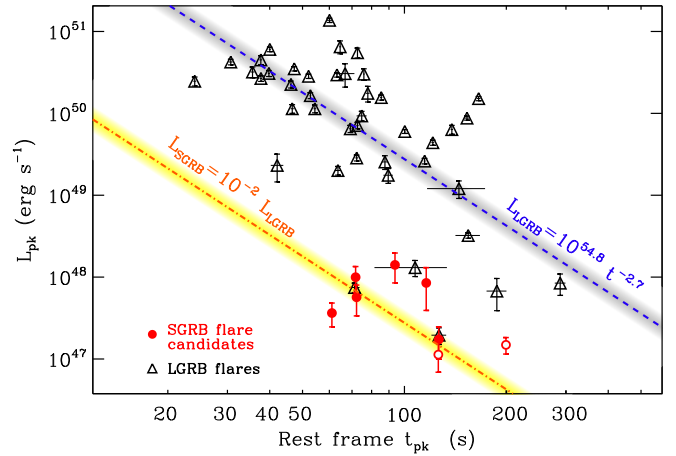}
      \caption{0.3-10 keV peak luminosity evolution with time for LGRB flares (black open triangles, from Chincarini et al. 2010) 
      and SGRB flare candidates with and without extended emission (red open and filled circles,
      respectively). Blue dashed line: best fitting power-law model for LGRB flares: 
      $L_{\rm{pk,F}}^{\rm{LGRB}}=10^{54.8\pm 0.4}\Big(\frac{t_{\rm{pk}}}{1+z}\Big )^{-2.7\pm0.5}$ and extrinsic scatter $\sigma=0.73\pm0.08$.
      Orange dot-dashed line: best fitting $L_{\rm{pk,F}}^{\rm{LGRB}}$ decay re-normalised by a factor 100 to match the
      observed SGRB flare candidates $L_{\rm{pk,F}}^{\rm{SGRB}}$.}
\label{Fig:Ltpk}
\end{figure}

\begin{figure}
\vskip -0.0 true cm
\centering
    \includegraphics[scale=0.40]{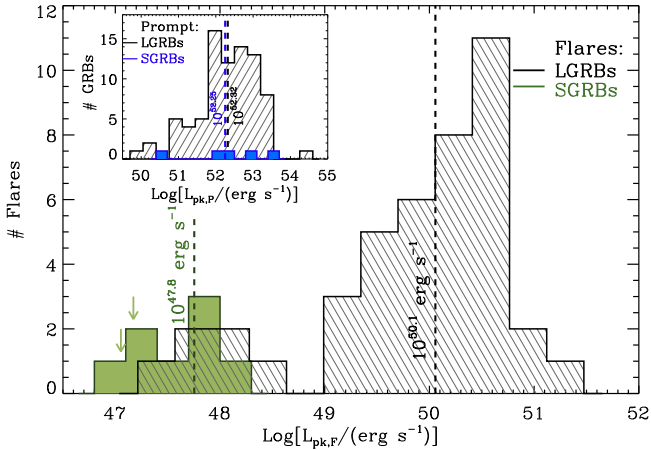}
      \caption{0.3-10 keV (observer frame) isotropic equivalent peak luminosity 
      $L_{\rm{pk,F}}^{\rm{LGRB}}$ of LGRB flares from Chincarini et al., 2010 (hatched histogram)
      compared to SGRB flare candidates (filled histogram); two vertical arrows mark the position of flare candidates in SGRBs with 
      extended emission. The vertical dashed lines mark the median values of the two distributions: 
      $L_{\rm{pk,F}}^{\rm{LGRB}}\sim 10^{50.1}\,\rm{erg\,s^{-1}}$; $L_{\rm{pk,F}}^{\rm{SGRB}}\sim 10^{47.8}\,\rm{erg\,s^{-1}}$.
      \emph{Inset:} 1-10000 keV rest frame isotropic equivalent peak luminosity distribution of the long (Nava et al., 2008) and 
      short GRBs prompt emission (Ghirlanda et al., 2009; 2010), with median value:  $L_{\rm{pk,P}}\sim10^{52.3}\,\rm{erg\,s^{-1}}$.}
\label{Fig:Lpiso}
\end{figure}

\begin{figure}
\vskip -0.0 true cm
\centering
    \includegraphics[scale=0.40]{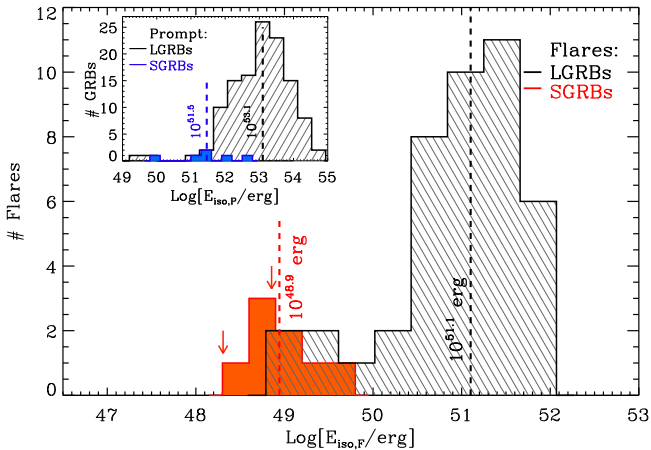}
      \caption{0.3-10 keV (observer frame) isotropic equivalent energy $E_{\rm{iso,F}}^{\rm{LGRB}}$ of LGRB flares 
      from Chincarini et al., 2010 (hatched histogram)
      compared to SGRB flare candidates (filled histogram); two vertical arrows mark the position of flare candidates in SGRBs with 
      extended emission.
       Vertical dashed lines: median $E_{\rm{iso,F}}^{\rm{LGRB}}\sim 10^{51}\,\rm{erg}$
      and $E_{\rm{iso,F}}^{\rm{SGRB}}\sim 10^{49}\,\rm{erg}$ values. \emph{Inset:} 1-10000 keV rest frame
      isotropic equivalent energy distribution for the prompt emission of the widest samples of long (hatched histogram) and short GRBs 
      (filled histogram) with firm spectral parameter estimates at the time of writing (Amati et al., in prep.), with median values:
      $E_{\rm{iso,P}}^{\rm{LGRB}}\sim10^{53.1}\,\rm{erg}$; $E_{\rm{iso,P}}^{\rm{SGRB}}\sim10^{51.5}\,\rm{erg}$.}
\label{Fig:Eiso}
\end{figure}

Count-rate light-curves have been converted into flux and luminosity (when possible) curves using 
the spectral information derived from a time-resolved spectral analysis where the spectral evolution 
of the source, if present, is properly accounted for. This procedure allows us to convert the best fitting
peak count rates of the X-ray flare candidates ($A$ parameter of Table \ref{Tab:flarepar}) into 0.3-10 keV
peak luminosities $L_{\rm{pk,F}}^{\rm{SGRB}}$ when the redshift of the source is known.   
Figure \ref{Fig:Ltpk} shows the evolution of the SGRB flare candidate $L_{\rm{pk,F}}^{\rm{SGRB}}$ with time compared
to the results obtained for LGRB flares by \cite{Chincarini10}; a comparison of the two 
distributions can be found in Fig. \ref{Fig:Lpiso}.
The isotropic equivalent energy $E_{\rm{iso,F}}^{\rm{SGRB}}$ has been determined integrating the best-fitting \cite{Norris05} 
luminosity-calibrated profiles from the onset time ($t_{s}$) to $t_{s}+100 w$ (where $w$ is the 
flare candidate width). The uncertainty arising from the spectral calibration has been 
propagated following standard practice into the final $L_{\rm{pk,F}}^{\rm{SGRB}}$ and $E_{\rm{iso,F}}^{\rm{SGRB}}$
 uncertainties listed in Tables \ref{Tab:sample} and \ref{Tab:flarepar}. We refer the reader to \cite{Margutti10a} for details on the 
light-curves and spectra extraction. Figure \ref{Fig:Eiso} shows the SGRB flare candidates
$E_{\rm{iso,F}}^{\rm{SGRB}}$ distribution compared to the values determined for LGRB flares, as 
computed by \cite{Chincarini10}.  
\subsection{Spectral time-lag computation}
\label{SubSec:lagcomputation}

The spectral lag is the time difference between the arrival of high-energy and low-energy photons.
For each GRB, the X-ray and $\gamma$-ray spectral lags ($\tau_{\rm{lag}}^{x}$ and $\tau_{\rm{lag}}^{\gamma}$,
respectively) and associated uncertainties have been determined using a  cross-correlation 
function (CCF) analysis. The CCF analysis requires the observations to have a fractional exposure equal to 1: 
this requirement excludes most of XRT observations taken in photon counting (PC) 
mode\footnote{\emph{Swift}-XRT automatically switches to the PC observing mode for count-rates below a few 
$\rm{count}\,\rm{s}^{-1}$ to minimise the presence of pile-up. In PC mode, it is not uncommon to have short time
intervals of no observation even during a single orbit. While the light-curve and spectra extraction procedures are 
basically insensitive to these short pauses, the CCF analysis would give un-reliable results.}. Among these,
the late-time ($t\sim 5\times10^{4}\,\rm{s}$) re-brightening of GRB\,050724 (Fig. \ref{Fig:lc050724051210}).
We closely follow the prescriptions by \cite{Stamatikos09} and \cite{Ukwatta10} for the CCF computation:
in particular, each CCF peak has been fitted using a third order polynomial; the number of points to be fitted
around the CCF peak has been allowed to vary from case to case with the possibility to specify asymmetric
intervals around the peak. In our analysis, a positive spectral lag is obtained if high energy photons lead
low energy photons.

The lag extraction is sensitive to a number of parameters: energy band pass of each comparative light-curve, 
temporal bin resolution, signal-to-noise ratio and presence of background emission
(i.e. in X-rays, the smooth X-ray decay underlying the time-variable signal). For the prompt $\gamma$-ray 
phase, the lag has been calculated using 4 ms light-curves (Section \ref{SubSec:BAT}) in the 50-100 keV and
100-200 keV energy bands. Time intervals covered by extended emission have been excluded. This 
allows us to perform a direct comparison with the time-lag values obtained for LGRBs observed by BAT
(\citealt{Ukwatta10}). Results are listed in Table \ref{Tab:sample}, column 5: the 8 SGRBs exhibit negligible 
$\tau_{\rm{lag}}^{\gamma}$.
 
\begin{figure*}
\vskip -0.0 true cm
\centering
    \includegraphics[scale=0.65]{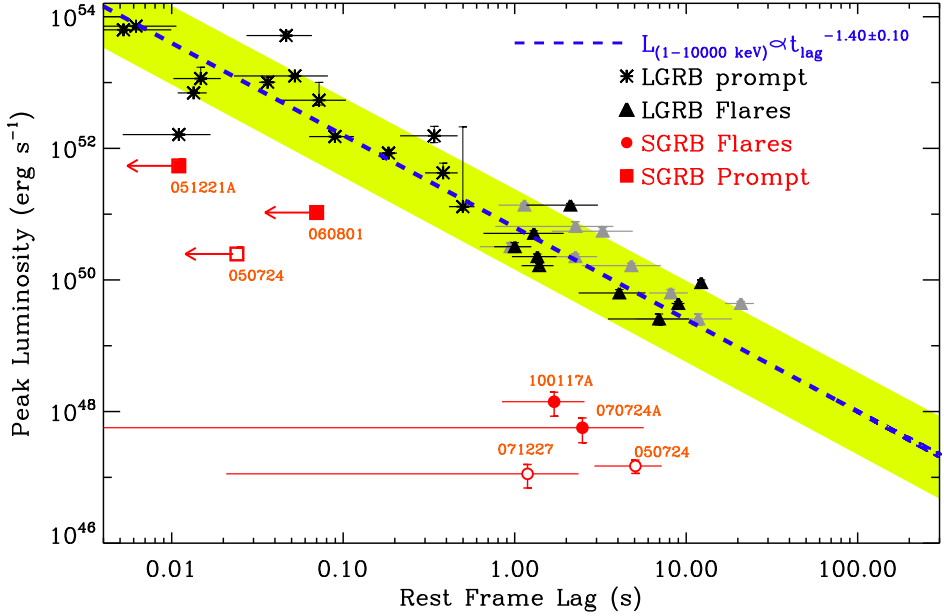}
      \caption{Lag-Luminosity plot. Red circles: CCF lag for candidate flares of SGRBs:
      open symbols refer to SGRBs with detected EE; black (grey) triangles: CCF (pulse peak)
      lag for  the sample of 9 flares of \emph{long} GRBs of Margutti et al., 2010b,  their Fig. 15. 
      Black stars: prompt $\gamma$-ray data from the gold and silver sample of Ukwatta et al., 2010. Red squares: $3\,\sigma$ upper limits
      to the prompt lag of SGRBs for which it is possible to estimate the peak luminosity: open symbols refer to SGRBs with detected EE.  
      The isotropic peak luminosity is computed in the $1-10^{4}\,\rm{keV}$ and $0.3-10\,\rm{keV}$ bands for prompt data and X-ray flares, respectively;
      the lag corresponds to the time difference between light-curve structures in 
      the 50-100 keV and 100-200 keV channels (prompt data) and 0.3-1 keV and 3-10 keV channels (X-ray flares). Blue dashed line:
      best fitting law for the LGRB \emph{prompt} emission data.}
\label{Fig:LagLumFlares}
\end{figure*}
  
In the X-rays the situation is complicated by the presence of a smoothly declining afterglow emission underlying 
the episodes of possible activity (see e.g. black solid line of Fig. \ref{Fig:lc100117A}).  Choices of 
re-binning time-scales, energy bands and/or temporal intervals giving origin to correlation values (CCF peak) $< 0.4$ 
have been discarded.  The choice of the energy bands to be compared is limited
by the XRT $0.3-10\,\rm{keV}$ coverage. For each SGRB, the X-ray time lag $\tau_{\rm{lag}}^{x}$ has been 
computed for different energy bands, giving consistent results: the $0.3-1\,\rm{keV}$ and  $3-10\,\rm{keV}$ 
energy bands have been finally chosen to perform a one -to -one comparison to the results obtained by 
\cite{Margutti10b} for flares detected in LGRBs. 
To this end, the LGRB flare time lags from  \cite{Margutti10b} have been re-calculated using the CCF analysis 
above (black dots in Fig. \ref{Fig:LagLumFlares}): in \citealt{Margutti10b} a \emph{pulse peak} lag 
was instead calculated (grey dots  in Fig. \ref{Fig:LagLumFlares}). The pulse peak lag 
is defined as $\tau_{\rm{lag}}^{\rm{peak}}\equiv t_{\rm{peak}}^{\rm{I}}-t_{\rm{peak}}^{\rm{II}}$ where $t_{\rm{peak}}^{\rm{I}}$
and $t_{\rm{peak}}^{\rm{II}}$ are the peak times of the best fitting profiles in the energy bands I and II, respectively.
As such,  $\tau_{\rm{lag}}^{\rm{peak}}$ is sensitive to the assumed pulse fitting model: 
while the dependency is limited in cases of bright events, the limited statistics of the SGRB X-ray light-curves
would cause the pulse \emph{peak} lag computation to be inaccurate. For this reason we refer to the \emph{CCF time lag}
for both short and long GRB data, in the $\gamma$-ray and X-ray regimes.
The light-curve time binning can potentially affect the derived $\tau_{\rm{lag}}^{x}$: for each SGRB the lag has been
computed on light-curve pairs with 10 different time binnings spanning the range $0.2-20$ s. The optimal time 
binning is defined as the lowest time scale giving origin to a CCF peak $>0.4$ and is listed in 
Table \ref{Tab:sample}. Larger binning time scales have been checked to produce consistent lag results. 
The window of time of investigation ($t_i$ and $t_f$ of Table \ref{Tab:sample}) has been determined selecting the time 
interval containing positive, at least $1-\sigma$ significant fluctuations around the smooth X-ray continuum 
(see \citealt{Margutti11} for details). 

For each SGRB, $t_i$ and $t_f$ have been varied of $\sim20\%$
both towards larger and lower values: consistent time lag values have been found. The sensitivity of the lag
measurement to the smooth X-ray light-curve decay underlying the candidate flares has been investigated calculating the lag 
$\tau_{\rm{lag}}^{x,\rm{sub}}$ on light-curve pairs where the contribution of the smooth afterglow component 
has been properly subtracted and uncertainties propagated  following the prescriptions by \cite{Margutti11}. 
For each SGRB  this procedure has led to consistent $\tau_{\rm{lag}}^{x,\rm{sub}}-\tau_{\rm{lag}}^{x}$ values
($\tau_{\rm{lag}}^{x,\rm{sub}}$ systematically has larger uncertainties due to the lower signal-to-noise of the subtracted 
light-curves). For this reason we refer to $\tau_{\rm{lag}}^{x}$ hereafter. Finally, we have tested and verified the 
robustness of our choice of energy bands to be compared, re-binning times and window
of time of investigation, by performing a number of simulations where artificial lags have been first introduced into 
the light curves and then successfully recovered. Results are reported in Table \ref{Tab:sample} .
Figure \ref{Fig:LagLumFlares} combines the $\tau_{\rm{lag}}^{x}$ and $L_{\rm{pk}}$ luminosity information in
the lag-luminosity plane and clearly shows that SGRB flare candidates fall off the prediction based on LGRB
flares and prompt pulses.

\section{results}
\label{Sec:Res}

\begin{figure*}
\vskip -0.0 true cm
\centering
    \includegraphics[scale=0.8]{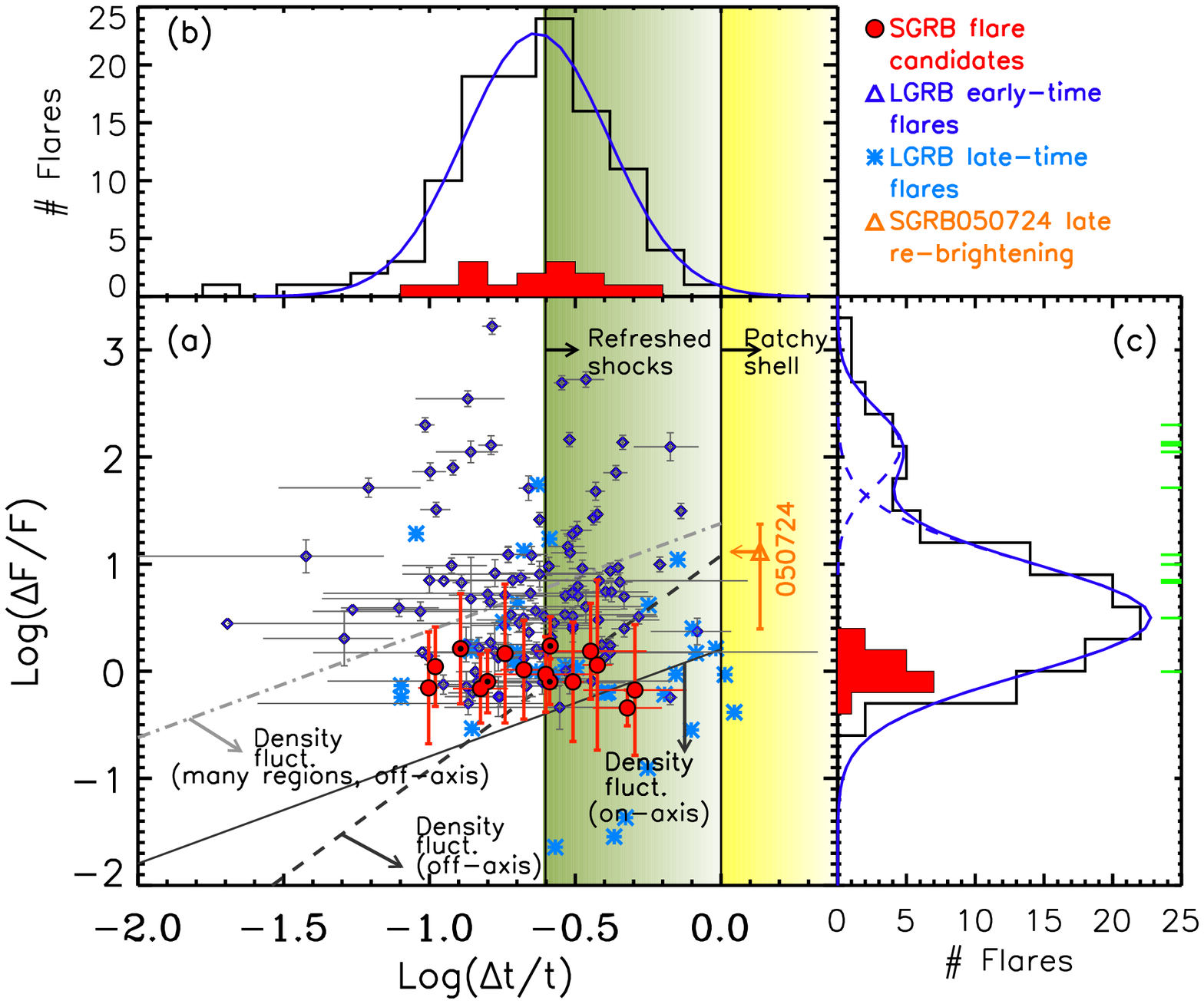}
      \caption{\emph{Panel (a):} Relative variability flux $\Delta F/F$ vs. relative variability time-scale $\Delta t/t\equiv w/t_{\rm{pk}}$ for the sample
      of X-ray flare candidates in SGRBs (filled circles), compared to early and late time LGRB X-ray flares (blue open diamonds
      and light-blue stars, respectively) from Chincarini et al., 2010 and Bernardini et al., 2011. The late time re-brightening detected in GRB\,050724
      is also shown for completeness with an orange open triangle (Bernardini et al., 2011). A small black dot marks  data coming from
      SGRBs with detected extended emission.
      Solid, dashed and dot-dashed lines mark the kinematically allowed regions
      in different scenarios according to Ioka et al., 2005, their  equations (7) and (A2). The $\Delta t/t$ and $\Delta F/F$ distributions 
      are portrayed in \emph{panels (b)} and \emph{(c)} adopting the same colour coding. The green tick marks in panel $(c)$ show the 
	flux contrast for the sub-sample of LGRB flares of Fig. \ref{Fig:LagLumFlares}.} 
\label{Fig:iokaShort}
\end{figure*}

\begin{table*}
\caption{X-ray and $\gamma$-ray properties for the sample of SGRBs analysed in this work. From left to right: GRB name: a ($^{*}$) 
	indicates an early-type host galaxy morphology (Fong et al., 2010; 2010b), while GRBs with detected extended emission are
	in boldface; redshift, duration and 
	average spectral photon index of the prompt 15-150 keV emission as determined from GCNs; 
	extended emission (EE) duration and EE to IPC (Initial Pulse Complex) intensity ratio ($R_{\rm{int}}$) from Norris et al. 2010a: upper limits on
	$R_{\rm{int}}$ are listed when no EE has been detected (EE with zero duration);
	$ t_{i}$ and $t_{f}$ define the temporal window for the X-ray lag calculation while the
	$\Delta t_{\rm{reb}}$ column reports the time scale used to re-bin the X-ray light-curve pairs; CCF time lag computed between $0.3-1\,\rm{keV}$ and
	$3-10\,\rm{keV}$; $0.3-10\,\rm{keV}$ isotropic equivalent peak luminosity in the time interval $t_i-t_f$ as determined from the Norris 2005 profile fit
	(see Table \ref{Tab:flarepar}); short lived (SL) or long lived (LL) X-ray afterglow according to the 
	classification by Sakamoto \& Gehrels, 2009.}
\begin{center}
\begin{tabular}{lllllllllllll}
\hline
GRB& z& $T_{90}$& $\Gamma_{\gamma}$&EE &$R_{\rm{int}}$ &$t_{i}$&$t_{f}$& $\Delta t_{\rm{reb}}$&$\tau_{\rm{lag}}^{x}$& $L_{\rm{pk,F}}^{\rm{SGRB}}$ & X-ray\\
       &   &  (s)       &  & (s)  &   & (s)     &      (s)                        &(s)                & (s) & ($10^{47}\,\rm{erg/s}$)  &afterglow  \\
\hline
\textbf{050724}$^{*}$   &0.258  & 3.00     &$1.71\pm0.16$&   104.4  & 0.0117  &213.8	&340.4		&3.00		&$6.4    \pm 2.7$ &  $1.49\pm0.34$ & LL\\
051210   &--        & 1.30	  &$1.10\pm0.30$&	0    &   0.0139&	87.3    	&171.0		&10.0		&$5.5    \pm 1.9$ &	 --                    & SL\\
\textbf{051227}   &--        & 8.00	  &$1.31\pm0.22$&	119.1 & 0.0540& 	101.2    	&177.2 		&4.00		&$-4.5   \pm7.2 $ &	 --                    & LL\\ 
060313   &--        & 0.74    &$0.71\pm0.07$&	0  	&   0.0005&  110.6    	&250.00		&8.00		&$30.5   \pm25.4$&	 --                    & LL\\
070724A &0.457  & 0.40	 &$1.81\pm0.33$&	0 	&    0.0074& 73.0    	&126.0  		&4.00		&$3.6    \pm4.6  $ &	 $5.68\pm2.32$& LL\\
\textbf{071227}   &0.383  & 1.80	  &$0.99\pm0.22$&    106.6 &0.0356 & 	126.18    	&201.0		&3.00		&$1.6    \pm1.6  $ &	 $1.13\pm0.44$& LL\\
090607   &--	   & 2.30	  &$1.25\pm0.30$&   0   	&  0.0016 & 76.1     	&173.3		&10.0		&$3.6    \pm 10.4$&	 --                    & SL\\  
100117A$^{*}$ &0.920   & 0.30	  &$0.88\pm0.22$&   0  & 	0.0030   & 86.1     	&238.5		&5.00		&$3.3    \pm1.6 $  &	 $14.09\pm5.60$& SL\\  
\hline
\end{tabular}
\end{center}
\label{Tab:sample}
\end{table*}

The data analysis of the previous sections leads to the following results:
\begin{itemize}
\item SGRB flare candidates appear both in early-type and late-type host galaxy environments, irrespective of the  
	short-lived (SL) or long-lived (LL) nature of the X-ray afterglow. 
\item Both SGRBs with a bright extended emission (EE) and
	SGRBs which lack this component show cases of statistically significant fluctuations superimposed over smoothly 
	decaying X-ray light-curves (Table \ref{Tab:sample}).
\item Flares in LGRBs are known to show a spectral hardening during the rise time and a softening during the decay time, reminiscent of the
	prompt emission (e.g. \citealt{Margutti10b} and references therein): as a result the hardness ratio (HR) evolution mimics the flare profile (see e.g.
	\citealt{Goad07}, their Fig. 9) while the spectral photon index evolution anti-correlates with the flare flux. In spite of the lower statistics and limited
	$\Delta F/F$ of SGRB flare candidates (Fig. \ref{Fig:iokaShort}), we find in the case with the best statistics a hint for correlation 
	between the HR evolution and the temporal profile of GRB\,101117A, with a photon index evolution that anti-correlates with the
	flux of the flare candidates (Fig. \ref{Fig:lc100117A}, middle and lower panels).
	In the other cases (Fig. \ref{Fig:lc050724051210}, \ref{Fig:lc051227060313}, \ref{Fig:lclast}, \ref{Fig:lclastlast}) the limited statistics 
	prevents us from drawing firm conclusions.
\item The SGRB flare candidates width evolution is roughly linear in time and consistent with the $w/(1+z)$  vs. $t_{\rm{pk}}/(1+z)$ relation established by LGRB
	flares over $\sim4$ decades in time (Fig. \ref{Fig:wtp}). The best-fitting law reads: $w/(1+z)=10^{-1.0\pm0.5}(t_{\rm{pk}}/(1+z))^{1.2\pm0.2}$.
	It is remarkable that data coming from LGRB flares both at early and very late time (beyond $t_{\rm{pk}}/(1+z)\sim 10^{5}\,\rm{s}$, Fig. \ref{Fig:wtp}, inset) 
	as well as temporal fluctuations in completely different systems like SGRB afterglows are consistent at zero-order with the same, 
	approximately linear, law. We refer to \citealt{Bernardini11} for a discussion of the possible bias affecting the $w$ vs. $t_{\rm{pk}}$
	relation.
\item SGRB flare candidates are  $\sim100$ times dimmer than LGRB flares at the same rest frame time (Fig. \ref{Fig:Ltpk}). 
	Selecting the sub-sample of LGRB flares detected in the same rest frame time interval $60\,\rm{s} \lesssim t_{\rm{pk}}/(1+z)\lesssim 250\,\rm{s}$
	of SGRB flare candidates, we obtain a median $\langle L_{\rm{pk,F}}^{\rm{LGRB}} \rangle \sim 10^{49.8}\rm{erg\,s^{-1}}$ to be compared to 
	$\langle L_{\rm{pk,F}}^{\rm{SGRB}} \rangle \sim 10^{47.8}\rm{erg\,s^{-1}}$ of the SGRB sample showed in Fig. \ref{Fig:Lpiso}.
	As a result, SGRB flare candidates fall off of a factor $\sim100$ the peak luminosity vs. time relation established by LGRB flares which reads: 
	$L_{\rm{pk,F}}^{\rm{LGRB}}=10^{54.8\pm0.4}(t_{\rm{pk}}/(1+z))^{-2.7\pm0.5}$. 
\item Short and long GRBs show a comparable $1-10^4\,\rm{keV}$ (rest frame) isotropic peak luminosity during their prompt emission, with a median 
	$\langle L_{\rm{pk,P}}\rangle \sim 10^{52.3}\,\rm{erg\,s^{-1}}$ (Fig. \ref{Fig:Lpiso}, inset).  On the contrary, the peak luminosity
	of flares of both categories evaluated at the same rest frame time $60\,\rm{s}<t_{\rm{pk}}/(1+z)<200\,\rm{s}$
	differ of a factor $\sim100$ as noted above. While for LGRBs the typical prompt 
	($1-10^4\,\rm{keV}$) to flare (0.3-10 keV) peak luminosity ratio $(L^{\rm{LGRB}}_{\rm{pk,P}}/L^{\rm{LGRB}}_{\rm{pk,F}})\sim 300$, 
	for SGRBs the same quantity reads: $(L^{\rm{SGRB}}_{\rm{pk,P}}/L^{\rm{SGRB}}_{\rm{pk,F}})\sim 3\times10^{4}$.
	Flare candidates in SGRBs are therefore \emph{less luminous} than expected using the prompt-to-flare luminosity scaling observed in LGRBs
	at the same flare rest frame time\footnote{The prompt emission peak luminosity is likely to be biassed towards the bright end
	of the real $L_{\rm{pk,P}}$ distribution, since a minimum signal-to-noise is needed to constrain the spectral parameters and calculate $L_{\rm{pk,P}}$
	in the 1-10$^4$ keV range: 
	this requirement is more severe in the case of SGRBs whose observed emission is usually less bright than LGRBs. However, Fig. \ref{Fig:Lpktpkrenorm}
	shows that the main conclusion of this paragraph remains true even after relaxing the requirement above.}.
\item SGRB flare candidates are $\sim100$ times \emph{less energetic} than LGRB flares (Fig. \ref{Fig:Eiso}), with a median 0.3-10 keV
	energy  $\langle E_{\rm{iso,F}}^{\rm{SGRB}}\rangle\sim10^{48.9}\,\rm{erg}$ 
	($\langle E_{\rm{iso,F}}^{\rm{LGRB}}\rangle\sim10^{50.9}\,\rm{erg}$). Since the width of LGRB flares and SGRB flare 
	candidates are comparable when evaluated at the same $t_{\rm{pk}}/(1+z)$, this result is a natural consequence of the 
	$\langle L_{\rm{pk,F}}^{\rm{LGRB}} \rangle/\langle L_{\rm{pk,F}}^{\rm{SGRB}} \rangle\sim 100$  reported above.
	On average, in the time interval $60\,\rm{s}\lesssim t_{\rm{pk}}/(1+z)\lesssim 250\,\rm{s}$ a flare in a LGRB emits $\sim0.6$\% of the 
	$1-10^4$ keV prompt $E_{\rm{iso,P}}^{\rm{LGRB}}$; the observed SGRB flare candidates isotropic energy is $\sim0.2$\% 
	the $1-10^4$ keV prompt $E_{\rm{iso,P}}^{\rm{SGRB}}$
	\footnote{Again, this ratio is likely to be a lower limit to real value due to the bias affecting the the sample of SGRBs with prompt $E_{\rm{iso}}$ 
	measure discussed in the previous paragraph.}. 
\item Flares and prompt pulses in LGRBs define a spectral lag-peak luminosity relation (Fig. \ref{Fig:LagLumFlares}): this finding is highly suggestive 
	of a common origin (\citealt{Margutti10b})\footnote{Note that flare and prompt lags are calculated in different energy bands and are not directly 
	comparable. The fact that flares define a lag-luminosity relation with slope very similar to the prompt data is however suggestive of a 
	strict connection between flares and prompt pulses. See \citealt{Margutti10b} for details.}. 
	On the contrary, SGRB prompt pulses are known to exhibit much shorter lags than expected if they were to follow the LGRB prompt pulses
	lag-luminosity relation (e.g. \citealt{Gehrels06}). Figure \ref{Fig:LagLumFlares} extends this behaviour to their flare candidates: like SGRB
	prompt pulses, flare candidates in SGRBs  \emph{fall off} the lag-luminosity relation defined by LGRBs.
\item Flares candidates of SGRBs in the  $\Delta F/F$ vs. relative variability time-scale ($\Delta t/t \equiv w/t_{\rm{pk}}$) 
	plane are compatible with variability arising from density fluctuations of many regions viewed off-axis: on the contrary, neither the refreshed-shock 
	nor the patchy-shell scenario is able to account for the observed properties of the entire sample (see \citealt{Ioka2005} for details on the
	definition of the various scenarios)\footnote{The smoking gun against a refreshed shock scenario would be the detection of a spectral 
	change contemporaneous to the flare candidates: while the statistics of the XRT light-curves of SGRBs is limited, in the case with best
	statistics the HR evolution is correlated to the candidate flare profiles and the continuum after the flaring emission is softer than the 
	emission detected during the period of temporal variability (Fig. \ref{Fig:lc100117A}). These findings favour an alternative explanation.}. 
	In particular: a K-S test comparing the $\Delta t/t$ distributions of LGRB flares and SGRB flare candidates
	reveals that they belong to the \emph{same} parent population at $\sim10$\% level of probability; the probability reaches the $88$\% level if 
	LGRB flares are selected in the SGRB peak time range  ($60\,\rm{s}\lesssim t_{\rm{pk}}/(1+z)\lesssim250\,\rm{s}$).
	This is consistent with the common $w/(1+z)$ vs. $t_{\rm{pk}}/(1+z)$ relation followed by short and long GRBs discussed above.
	On the contrary, no X-ray flare candidate in a SGRB shows a relative variability flux $\Delta F/F > 2$ in strong \emph{contrast} with the 
	LGRB $\Delta F/F$ distribution at comparable $t_{\rm{pk}}/(1+z)$ (Fig. \ref{Fig:iokaShort}). Such prominent flares would be
	easier to detect, so that it is unlikely that an observational bias could explain the present lack of detection. 
	A K-S test comparing the two distributions shows that the probability that 
	LGRB and SGRB flare candidates share the same $\Delta F/F$ parent population is as low as $3.3\times10^{-4}$. This result partially inherits 
	the uncertainty affecting the completeness of both distributions for very small $\Delta F/F$ values. Another source of uncertainty 
	arises from the difficulty in evaluating the continuum underlying the possible flare emission  in SGRBs when data are particularly sparse.
	In spite of these limitations, after more than 6 years of \emph{Swift} observations  (and $\sim60$ SGRB afterglows detected)
	there is still no SGRB showing a prominent ($\Delta F/F>10$) fast-varaibility $\Delta t/t \ll1$ feature during its X-ray afterglow.  
	The SGRB flare candidate $\Delta F/F$  is instead more similar to the relative variability flux of flares in LGRBs detected at \emph{late} times 
	($t_{\rm{pk}}>1\,\rm{ks}$, light-blue stars in Fig. \ref{Fig:LagLumFlares}, main panel; \citealt{Bernardini11}): the two $\Delta F/F$ 
	distributions share the same parent distribution at $\sim21$\% level of probability (K-S test).
\end{itemize}

The results above demonstrate the complexity characterising the  SGRB flare candidates phenomenology: 
Table \ref{Tab:diff} reports a summary of their properties when compared to LGRB flares. The main
result is that the population of SGRB X-ray flare candidates only \emph{partially} share the set of observational 
properties of LGRB X-ray flares detected at the same rest frame time: are there real X-ray flares in SGRBs?
A detailed discussion is provided below.

\begin{table}
\caption{Summary of the properties of SGRB flare candidates compared to LGRB X-ray flares observed at the
\emph{same}  $t_{\rm{pk}}/(1+z)$.}
\begin{center}
\begin{tabular}{lll}
\hline
\textbf{Property} & = & $\neq$ \\ 
 \hline
Width  & $w(t_{\rm{pk}})$  &\\
Relative variability &$\Delta t/t$         &\\
Relative variability flux & &$\Delta F/F$         \\
&&\\
Peak luminosity  & &$\frac{L_{\rm{pk,F}}^{\rm{SGRB}}}{L_{\rm{pk,F}}^{\rm{LGRB}}}\sim 0.01 $ \\
&& \\
Isotropic energy  & &$\frac{E_{\rm{iso,F}}^{\rm{SGRB}}}{E_{\rm{iso,F}}^{\rm{LGRB}}}\sim 0.01 $\\
Lag-luminosity             && $L_{\rm{pk}}(\tau_{\rm{lag}}^{x})$\\
Flare to prompt energy ratio &$\frac{E_{\rm{iso,F}}}{E_{\rm{iso,P}}}$ \\
&&\\
Flare to prompt luminosity ratio & &$L_{\rm{pk,F}}/L_{\rm{pk,P}}$ \\
Flare to prompt pulse width ratio && $w_{\rm{F}}/w_{\rm{P}}$\\
\hline
\end{tabular}
\end{center}
\label{Tab:diff}
\end{table}

\section{Discussion}
\label{Sec:Disc}
Observations show that like LGRBs, at least some SGRB X-ray afterglows deviate from 
a smooth power-law decay and show variability. In the following we discuss the properties of SGRB 
flare candidates providing a one-to-one comparison with LGRB X-ray flares. The aim is to better
understand the origin of the short burst afterglow variability and uncover potential links with
the prompt phase.

With $\Delta t/t>1$, the late-time ($t_{\rm{pk}}\sim5\times10^4\,\rm{s}$) re-brightening of 
GRB\,050724 strongly differs from the properties of the entire sample of flare 
candidates detected in SGRBs thus questioning its classification as flare-like episode
(\citealt{Panaitescu06}: see however \citealt{Grupe06}; \citealt{Campana06}; \citealt{Malesani07}). 
\cite{Malesani07} report the detection of an optical and radio re-brightning  associated
to the X-ray bump which is unusual if compared to the standard properties of 
X-ray flares, while being more common to late-time re-brightneings observed in LGRBs as well
(see e.g. GRB\,081028, \citealt{Margutti10a}). In addition, no hard-to-soft evolution can be
detected in the X-ray data (\citealt{Evans10}), which is instead typical of flares (\citealt{Goad07}) and
prompt pulses in LGRBs (\citealt{Hakkila11}).
In the following we focus our attention on SGRBs fast variability
($\Delta t/t< 1$) referring the reader to \cite{Bernardini11} for a complete discussion of the late-time
behaviour of GRB\,050724. 
\subsection{SGRB local and global environment}
\label{SubSec:environment}
The standard model (see \citealt{Nakar07} for a recent review) explains 
the X-ray afterglow of long and short GRBs as synchrotron radiation 
arising from the deceleration of a relativistic blast wave into the external medium.  If the 
shock front is homogeneous and expands into a smooth ambient density, a smooth
afterglow light-curve is expected. In this context, variability in the X-ray afterglow could be caused
by re-freshed shocks (i.e. shocks caused by slow shells catching up with the leading, decelerated
shell at late times, \citealt{Kumar00a}, \citealt{Granot03}):  Fig. \ref{Fig:iokaShort} shows that half of the SGRB flare 
candidates sample do not comply with this scenario\footnote{See however \cite{Granot03}; \cite{Vlasis11}:
sharp \emph{optical} and \emph{radio} flares could be produced by collision of ultra-relativistic shells.}. 
Furthermore, the spectral variability shown in Fig. \ref{Fig:lc100117A}, lower panel,  
makes it difficult to interpret the flare candidates in the re-freshed shocks scenario.

A first alternative is to relax the assumption
on the homogeneity of the shock front (\citealt{Kumar00b}): an intrinsic angular structure of the emitting surface is able
to produce variability with a characteristic time-scale $\Delta t \geq t$ if the angular fluctuation is
persistent (patchy shell model, \citealt{Nakar04})\footnote{Details on variability arising from a 
time-varying anisotropic emitting surface can be found in \cite{Ioka2005}.}: no SGRB flare candidate
is consistent with this expectation (Fig. \ref{Fig:iokaShort}, main panel). 

A second alternative  invokes the presence of ambient density fluctuations either caused by turbulence in the ISM or
by variable winds from the progenitor. 
From Table \ref{Tab:sample} it is however apparent that temporal X-ray variability has been detected
for SGRBs residing both in early-type and late-type host galaxies which likely have very different ISM properties.
In particular GRB\,050724 and GRB\,100117A are the unique two SGRBs with secure early-type host galaxy association 
(\citealt{Barthelmy05b}; \citealt{Fong2010}). This suggests that  the ISM turbulence
is unlikely to provide a feasible physical mechanism for the detected variability. Note however that the limited size of  our
sample prevents us from quantitatively discussing the correlation between the appearance of
flare candidates and host environment.

Different progenitor models of SGRBs lead to distinct predictions on their \emph{local} environment as well. In 
particular, according to the standard compact binary merger scenario (NS-NS or NS-BH, \citealt{Eichler89};
\citealt{Narayan92}), no variable wind is expected from the progenitor.  
An origin of SGRB flare candidates from density fluctuations of the  circumburst environment is therefore 
considered unlikely. Alternative scenarios leading to SGRB environments with circumburst densities comparable
to LGRBs have been however explored by \cite{Nysewander09} to explain the similar $F_{R}/F_{X}$ ratio (where
$F_R$ and $F_X$ stand for the afterglow flux density at 11 hrs post-trigger in the $R$ and $X$-ray bands, respectively).
A systematic difference between sub-galactic environments able or unable to produce variability in the X-ray 
afterglow could in principle be revealed by different offsets  with 
respect to the host galaxy centers: however, the present sample of SGRBs with flare candidates 
includes both SGRB with large offsets (e.g. $\delta=14.80\pm0.34\,\rm{kpc}$ for SGRB\,071227, \citealt{Fong2010b};
\citealt{Davanzo2009}) and events with very small offsets (e.g. $\delta=0.47\pm0.31\,\rm{kpc}$, SGRB\,100117A, \citealt{Fong2010}).
While the observed offset distribution is incomplete, these data suggest that the properties of the   
\emph{local} environment of SGRBs are not the key parameters determining the presence of flare candidates
in their X-ray light-curves. This conclusion is strengthened by the results of Sect. \ref{SubSec:Energetics}.

Flares candidates and prompt pulses could alternatively share a common origin.
In this case, flare candidates would bring no information on the external medium density.
We refer to this interpretation as the \emph{"internal origin"} possibility (see Sect. \ref{Sec:intor}).
\subsection{Flare candidates and extended emission}
\label{SubSec:EE}
3 bursts in our sample (boldface in Table \ref{Tab:sample}) present evidence for extended 
emission (EE): a  long-lasting ($\Delta t\sim 10^2\rm{s}$) 
soft X-ray tail that follows the short hard spike in the prompt phase. 
\cite{Norris10a} analysed a sample of 51  \emph{Swift}-BAT SGRBs looking for the presence of EE
 in their $\gamma$-ray data and concluded that $\sim3/4$ of the BAT SGRBs are \emph{not} 
accompanied by an EE component.  
In particular, in this work it is shown that
the \emph{upper limit} on the EE to IPC (initial pulse complex) intensity ratio of the SGRBs of our 
sample (i.e. with flare candidates) not showing EE is a factor $\gtrsim10$ below the standard values found for GRBs with EE  
(\citealt{Norris10a}, their Fig. 1). This implies that in those cases the EE is either
very weak or absent. The inverse is also true: some SGRBs with bright EE are accompanied by a smooth X-ray
light-curve at $t\gtrsim 80$ s (see e. g. GRB\,080503, \citealt{Norris10a}, their Table 1). 

While it is still unclear if the 
origin of the EE component is related to the prompt emission, the afterglow or a third, unknown physical process 
(\citealt{Perley09}; \citealt{Norris10a}; \citealt{Norris11}), our analysis shows that the presence of a bright 
EE does not imply the presence of flare candidates and viceversa.
We stress that the higher average brightness characterising the
XRT light-curves of GRBs with EE with respect to those without EE, would naturally bias the result in the opposite 
direction (i.e. it would favour temporal fluctuations to be detected in XRT light-curves with the best statistics, leading to a 
biassed flare-EE connection) thus strengthening our conclusion. The limited size of our sample (which is however
the widest possible at the time of writing) does not allow
us to quantitatively discuss the flare candidates vs. EE correlation (or lack thereof).

Our analysis cannot however exclude that SGRB
flare candidates constitute temporal fluctuations superimposed on (and physically linked to) the X-ray tail of the EE, as
suggested by the epoch of flare candidates detection ($t_{\rm{pk}}/(1+z)\lesssim 200$ s). Furthermore,
the limited range spanned by the flare-to-continuum flux ratio $\Delta F/F$ (with $\Delta F/F\sim1$, Fig. \ref{Fig:iokaShort})
is suggestive of a physical link between flare candidates and the underlying continuum (i.e. the EE). 
Again, flare candidates would be associated to both bright and faint (undetected in the $\gamma$-rays) EEs.
\subsection{Time scales}
\label{SubSec:time}
The relative time-scale distribution $\Delta t/t$ of SGRB flare candidates (Fig. \ref{Fig:iokaShort}, upper panel) is compatible 
with being drawn from the \emph{same} parent population of flares detected in LGRBs at similar $t_{\rm{pk}}/(1+z)$ 
at $\sim88$\% level of probability:
with a median value of $\langle\Delta t/t\rangle=0.25$  and extending from 0.1 up to 0.5 it furthermore satisfies the limit $\Delta t/t \leq 1$
which is the classical argument against fluctuations arising from the external shock (e.g. \citealt{Zhang06}). 
The $\Delta t/t$ distribution therefore does not support an external shock origin
for flare candidates in SGRBs (see however \citealt{Dermer08}), 
in agreement with the findings of Sec. \ref{SubSec:environment}.

 The evolution of the flare duration with time $w(t_{\rm{pk}})$ for $t_{\rm{pk}}>T_{90}$ (Fig. \ref{Fig:wtp}) 
observationally distinguishes flares from prompt pulses in  LGRBs (\citealt{Margutti10b}): as 
time proceeds, LGRB flares becomes wider and wider following the best fitting law: 
$w/(1+z)\sim (t_{\rm{pk}}/(1+z))^{1.2\pm0.2}$. This quasi-linear regime establishes for 
$t_{\rm{pk}}/(1+z)>T_{90}/(1+z)$, and extends 4 decades in time 
up to $t_{\rm{pk}}/(1+z)\gtrsim10^5$ s. Notably, while the $E_{\rm{iso,P}}^{\rm{LGRB}}$ released during the LGRB prompt
emission spans more than 3 decades in energy (Fig. \ref{Fig:Eiso}), likely reflecting different properties of the LGRB 
central engines, the \emph{duration} of subsequent episodes of activity seems to follow a universal evolution with 
limited dependence on the initial energy budget. Flares with different amplitudes $A$, flux ratios $\Delta F/F$ and
fluences show similar $w/(1+z)$ when appearing at the same $t_{\rm{pk}}/(1+z)$  (\citealt{Chincarini10}), suggesting that 
the main parameter driving the flare width evolution in a LGRB is the elapsed time from the 
explosion $t_{\rm{pk}}/(1+z)$\footnote{As a second level of approximation, one should consider that more prominent
flares ($A\gg1$) tend to be wider (\citealt{Margutti10b}).}.

Figure \ref{Fig:wtp} shows that flare candidates in SGRBs are consistent with the \emph{same} quasi-linear 
temporal scaling: from the width measurement it is not possible to distinguish a flare in a LGRB from a flare
candidate in a SGRB. The temporal properties of the prompt emission of long and short GRBs are clearly 
different in terms of duration and variability (e.g. \citealt{Nakar02}): however, $\sim30$ s later, the width of
fluctuations superimposed over their X-ray light-curves seems to have lost any information on the nature of central engines
able to produce $\gamma$-ray photons with such different temporal properties.
In both cases the $w$ evolution is driven by the $t_{\rm{pk}}$, irrespective of their different initial conditions
(and initial variability time scales): while for flares $w_{\rm{F}}^{\rm{SGRB}}/(1+z)\sim w_{\rm{F}}^{\rm{LGRB}}/(1+z)$ 
at similar $t_{\rm{pk}}/(1+z)$, for prompt pulses $w_{\rm{P}}^{\rm{SGRB}}\ll w_{\rm{P}}^{\rm{LGRB}}$ (with 
$w_{\rm{P}}^{\rm{SGRB}}\sim 0.05$ s to be compared to $w_{\rm{P}}^{\rm{LGRB}}\sim1$ s, observer frame values,
\citealt{Nakar02}; \citealt{Nakar02b}\footnote{Note however that the first 1-2 s of bright LGRBs display 
variability time scales comparable to the SGRB prompt emission, \cite{Nakar02}.}).

This finding strongly suggests that the origin of the quasi-linear $w/(1+z)$ vs. $t_{\rm{pk}}/(1+z)$ scaling must be within
what is in common for the long and short GRB models, irrespective of the progenitors, environment, lifetime, variability 
time-scales and prompt energy release of their central engines. Both models require the prompt emission
to originate from a relativistic outflow  (see e.g. \citealt{Piran04}): if the longer and longer duration of flares is of dynamical origin and
dominated by the expansion of the emitting regions, no memory of the properties of the central engine
which launched the relativistic outflow would be preserved so that long and short GRBs would display 
flares with similar $w(t_{\rm{pk}})$. Alternatively, in the context of accretion models the $w(t_{\rm{pk}})$
relation originates from the viscous evolution of the hyper-accreting disk around the newly born black hole,
common ingredient of the likely progenitors of the two classes (\citealt{Perna06}). Our results would imply 
a strict analogy between the mechanisms that regulate the late-time evolution of accreting disks originating
from collapsars and mergers of compact objects, irrespective of their different masses and life times
(according to the standard scenario accreting disks related to SGRBs are likely to be  less massive then 
LGRB disks and short-lived, as suggested by the observed prompt duration. See \citealt{Nakar07} for details).

\cite{Lazzati11} recently suggested that instabilities arising from the propagation of the jet through the
disrupting star could explain the presence of flares in LGRBs with $\Delta F/F \lesssim 10$. Even
assuming a continuous and featureless release of energy by the central engine,
the pressure of the surrounding stellar material would provide the physical origin for the intermittent 
flare behaviour, naturally explaining the universal (i.e. with limited sensitivity on the star properties and
energy budget) quasi-linear $w(t_{\rm{pk}})$ relation (\citealt{Lazzati11}, their Eq. (6)).
However this model cannot account neither for presence of flares in SGRBs nor for
their similarity to LGRB flares in the $w/(1+z)$ vs. $t_{\rm{pk}}/(1+z)$ plane: according to the majority of 
SGRB progenitor scenarios the engine is "exposed" and there is no stellar material the jet could interact
with. As a consequence, if the $w(t_{\rm{pk}})$ relation in LGRBs originates from the interaction with the 
progenitor stellar material, it is difficult to explain why flare candidates in SGRBs, while originating from
a completely different physical scenario, are however consistent with 
the same relation, as observed. Our results therefore imply that either the LGRB flare $w(t_{\rm{pk}})$ 
relation does not originate from the jet-stellar material interaction or that the progenitors
of long and short GRBs are much more similar than previously thought  (see e.g. \citealt{Lazzati10}). 

Finally, our findings suggest that
while the variability time scales measured during the prompt emission could still directly reflect the intrinsic
variability of the central engine (see \citealt{Piran04} and references therein), absolute measures (i.e. not re-normalised) 
of flare widths likely do not
(if time-dilated by physical mechanisms which are only indirectly related to the central engine). 
On the contrary, the \emph{ratio} of  interesting time-scales of the same flare profile, being subject to the same temporal
stretching, could still retain an imprint of the original mechanism at work: this would explain why in LGRBs the flare 
rise time $t_{\rm{r}}$ and decay time $t_{\rm{d}}$, like the $w$,  linearly evolve with $t_{\rm{pk}}$ (\citealt{Chincarini10}; 
\citealt{Bernardini11}) preserving 
their ratio $t_{\rm{r}}\sim 0.5 t_{\rm{d}}$ over 4 decades in $t_{\rm{pk}}$ and leading to flares with asymmetry values
very similar to the prompt pulses (while being a factor $\geq100$ wider)\footnote{Note that given the limited statistics
of the SGRB X-ray afterglows, nothing can be said about the asymmetry of SGRB flare candidates.}.
\subsection{Flux contrast}
\label{SubSec:flux}

While the $w(t_{\rm{pk}})$ and $\Delta t/t$ measurements do not allow us to distinguish a flare in a LGRB from 
a flare candidate in a SGRB\footnote{The limited number of flare candidates plays a role in this statement. We cannot exclude that 
a significant improvement of the SGRB and LGRB statistics could lead to the detection of secondary effects.}, 
the flux contrast distributions $\Delta F/F$ of the two populations are significantly different 
(Fig. \ref{Fig:iokaShort}), with SGRB flare candidates having  systematically \emph{lower} $\Delta F/F$ values. 
In contrast to LGRBs, none of the $\sim 60$ \emph{Swift} X-ray afterglows 
associated to SGRBs shows cases of powerful  ($\Delta F/F \gg 5$), highly variable $\Delta t/t\ll1$ flares. 
A \emph{Chandra} observation of 9 X-ray photons from SGRB\,050709\footnote{GRB\,050709 is a short burst 
detected by \emph{HETE} for which \emph{Swift} did not
do the follow up. For this reason it is not included in the present sample. The episode of prominent 
variability is possibly related to a statistical fluctuation.} $\sim15$ days after the explosion 
led \cite{Fox05} to conclude the presence of high-amplitude ($\Delta F/F \approx 10$), fast variability 
($\Delta t/t \approx 0.01$) in its X-ray afterglow. \emph{Swift} observations suggest that this kind of variability, 
if present, is extremely rare. SGRBs lack the presence of prominent fast-rise exponential 
decay features superimposed over their X-ray afterglow for $t_{\rm{pk}}/(1+z)>60$ s.

\cite{Lazzati11} predicted that if flares in SGRBs originate from the intrinsic variability of their inner engine, 
their $\Delta F/F$ distribution should be analogous to the brighter population ($\Delta F/F\sim 100$) of LGRB 
flares. With a maximum $\Delta F/F\approx 2$ the detected SGRB flare candidates are \emph{not} consistent with 
these expectations and populate instead the \emph{low} end of the LGRB distribution unless the SGRB continuum flux $F$
has been overestimated by a factor $\sim100$ which we consider unlikely. An interesting possibility is that the 
X-ray light-curves of SGRBs are dominated by an emission component which is \emph{not} present in
the LGRB afterglows (such as the EE): this would lead to systematically lower $\Delta F/F$ for SGRBs when compared to LGRB
flares. The clustering of the $\Delta F/F$ ratio around 1 would suggest a link between flare candidates
and underlying continuum.
However,  the strong correlation found by \cite{Nysewander09}  between the X-ray afterglow flux and 
prompt fluence of \emph{both} long and short GRBs reveals a high degree of similarity between their 
X-ray afterglows at least at late times (i.e. 11 hrs, observer frame). A detailed comparison of short and long
GRB X-ray afterglows at \emph{early} times is in preparation.

The SGRB $\Delta F/F$ observed at $t_{\rm{pk}}\sim100\,\rm{s}$ is instead typical of LGRB flares detected
at much later times: $t_{\rm{pk}}\gtrsim 1\,\rm{ks}$ (\citealt{Bernardini11}). 
From Fig. \ref{Fig:dfevol} it is apparent that SGRB flare candidates are
consistent with the $\Delta F/F$ distribution of LGRB flares detected at the \emph{same} $t_{\rm{pk}}/T_{90}$
epoch (where $T_{90}$ is the duration of the prompt 15-150 keV emission): 
differences instead arise if we compare the properties of the two classes at the same
$t_{\rm{pk}}/(1+z)$.  While the $T_{90}$ is possibly a crude estimation of the intrinsic time 
scale of evolution of the central engine, this result opens the possibility that prominent fluctuations are not currently
detected in SGRB afterglows because of the intrinsically late-time \emph{Swift} re-pointings: $t_{\rm{rep}}\sim100$ s
corresponds to $t_{\rm{rep}}/T_{90}\sim100-1000$ (SGRBs) and to $t_{\rm{rep}}/T_{90}\sim1-10$ (LGRBs).
From another perspective this finding directly connects the flux properties of SGRB flare candidates to the 
evolution time scale of their central engines. 
The prompt
$T_{90}$ qualifies as a good proxy for the intrinsic time scale that drives the subsequent flux evolution of 
flares and flare candidates in long and short GRBs, respectively. This conclusion is strengthened by the
results of Sec. \ref{SubSec:Energetics}.

\begin{figure*}
\vskip -0.0 true cm
\centering
    \includegraphics[scale=0.6]{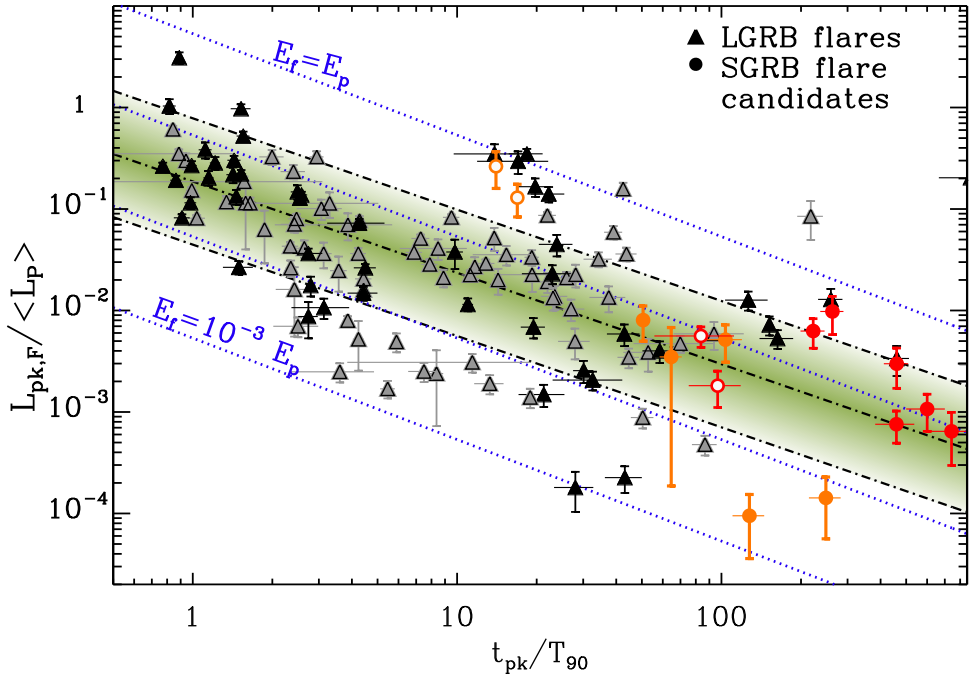}
      \caption{Re-normalized peak luminosity vs. re-normalized peak time for the sample of LGRB flares from Chincarini et al. (2010)  
      and Bernardini et al. (2011) (triangles)  with $t_{\rm{pk}}/T_{90}\lesssim10^3$ and SGRB flare candidates (circles).  
      Dark and light colours distinguish events with and without redshift measurement to allow a direct
      comparison with Fig. \ref{Fig:Ltpk}. Open symbols refer to SGRBs with detected extended emission.
      Black dot dashed line: best fitting power-law model for the LGRB sample: 
      $L_{\rm{pk,F}} / \langle L_{\rm{P}}\rangle=10^{-0.72\pm0.03}(t_{\rm{pk}}/T_{90})^{-0.9\pm0.1}$ and extrinsic scatter $\sigma=0.62\pm0.01$.
      The shaded area marks the $\pm 1\,\sigma$ region around the best fit. From top to bottom, the blue dotted lines mark the 
      $E_{\rm{iso,F}}/E_{\rm{iso,P}}=1,0.1,0.01, 0.001$ regions of the plane, where $E_{\rm{iso,F}}$ and $E_{\rm{iso,P}}$
      have been calculated in the 0.3-10 keV and 15-150 keV energy bands, respectively. The flare width vs. peak time relation from
      Chincarini et al. (2010) has been used.}
\label{Fig:Lpktpkrenorm}
\end{figure*}
\begin{figure}
\vskip -0.0 true cm
\centering
    \includegraphics[scale=0.37]{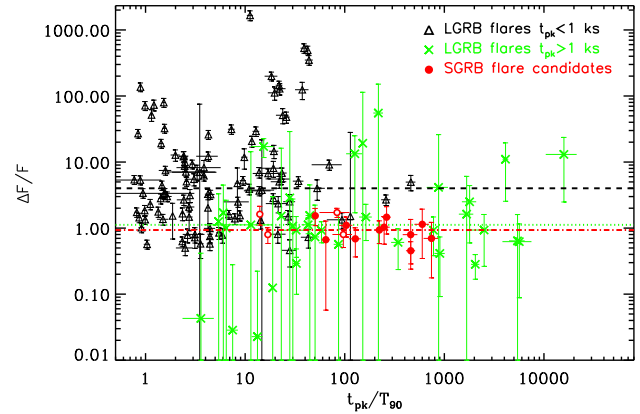}
      \caption{Relative variability flux $\Delta F/F$ evolution as a function of $t_{\rm{pk}}/T_{90}$: black triangles and green stars
      indicate LGRB flares with $t_{\rm{pk}}<1$ ks (Chincarini et al., 2010) and $t_{\rm{pk}}>1$ ks (Bernardini et al., 2011), 
      respectively. Red open and filled circles: flare candidates in SGRBs with and without extended emission, respectively.
      Horizontal dashed, dot and dot-dashed lines mark the $\Delta F/F$ median value for
      the three samples.}
\label{Fig:dfevol}
\end{figure}

\subsection{Energetics}
\label{SubSec:Energetics}
\subsubsection{Flare $L_{\rm{pk}}$ evolution with time} 
\label{SubSubSec:Energetics1}
While SGRB prompt pulses compete with those of LGRBs in terms of peak luminosity, the same is not true for
their late-time variability: SGRB flare candidates are a factor $\sim10^2$ dimmer that expected (Fig. \ref{Fig:Ltpk}). 
Sec. \ref{SubSec:flux} and Fig. \ref{Fig:Eiso} suggest that long and short GRBs might be astrophysical systems 
\emph{evolving} on completely different time-scales and with different energy budgets but otherwise 
based on a similar physical mechanism of emission. In that case, 
we would expect the evolution of the flare luminosity to exhibit much better agreement between long and short GRBs
once the intrinsic time-scale and central engine energy scaling are properly accounted for.
Figure \ref{Fig:Lpktpkrenorm} shows that this is indeed the case: long and short GRBs are consistent with a 
common (albeit highly scattered) 
$L_{\rm{pk,F}}/\langle L_{\rm{P}}\rangle\sim (t_{\rm{pk}}/T_{90})^{-0.9\pm 0.1}$ scaling  (where $\langle L_{\rm{P}}\rangle$ 
and $T_{90}$ are the average prompt luminosity and duration of each GRB in the 15-150 keV energy range,
respectively). Figure \ref{Fig:Lpktpkrenorm} therefore establishes a direct connection between the properties of SGRB flare candidates
and LGRB flares, providing further support to a common, likely \emph{internal} origin (see Section \ref{Sec:intor}).

In the context of a one to one comparison of short vs. long 
GRB X-ray variability, the re-normalised peak luminosity and re-normalised peak time are the most relevant quantities: 
the observed $T_{90}$ and $\langle L_{\rm{P}} \rangle$, with their sensitivity to the instrument threshold and energy band passes,
likely provide crude (but nevertheless the best) approximations to the exact values to be used
(while being partially responsible for the large scatter of the relation).
In particular the loss of total fluence in the prompt $\gamma$-ray due to the limited \emph{Swift}-BAT band-pass likely
affects short more than long GRBs. As a result the SGRB $L_{\rm{pk,F}}/\langle L_{\rm{P}}\rangle$ values may be 
overestimated when compared to LGRB values\footnote{This effects would mainly depend on the different prompt
emission spectral peak energy $E_{\rm{peak}}$ of long and short GRBs: however, \cite{Ghirlanda09} 
showed that SGRBs have a harder low-energy spectral component but only slightly higher $E_{\rm{peak}}$ when compared to LGRBs.}. 
However, Fig. \ref{Fig:Lpktpkrenorm} shows that even a factor of several 
of overestimation (\citealt{Nysewander09}) would not strongly affect our conclusions. 
\subsubsection{Implications of the $L_{\rm{pk,F}}/\langle L_{\rm{P}}\rangle$ vs. $t_{\rm{pk}}/T_{90}$ relation}
\label{SubSubSec:Energetics2}
SGRB flare candidates are consistent with the highly scattered $L_{\rm{pk,F}}/\langle L_{\rm{P}}\rangle$ vs. $t_{\rm{pk}}/T_{90}$ relation
established by LGRB flares: since the origin of LGRB flares is likely connected to their prompt pulses
 (\citealt{Krimm07}; \citealt{Margutti10b}), we speculate that a similar physical mechanism (except for its energy budget and life-time)
powers long and short GRBs: observationally speaking, the main distinction between flares and prompt 
emission in LGRBs is the evolution of the former with time for  $t_{\rm{pk}}>T_{90}$.
It is therefore natural to expect a similar scaling of LGRB flares and SGRB flare candidates in terms
of $t_{\rm{pk}}/T_{90}$ if they share a common origin.

A comparison of the prompt $\gamma$-ray emission of SGRBs to the initial 2 s of LGRBs reveals a high 
degree of similarity in the pulse duration distributions and correlations in the temporal structure of the 
two classes (\citealt{Nakar02}; \citealt{McBreen01})\footnote{The temporal evolution of pulses as a function 
of \emph{frequency} (i.e. the spectral lag) shows however dissimilarities, as discussed in Section \ref{SubSec:laglum}.}; 
an analogous study was performed by \cite{Ghirlanda09} in the spectral domain: based on the spectral analysis of the prompt emission 
of 79 short and 79 long GRBs detected by BATSE the authors showed that no difference is found comparing 
the spectral properties of SGRBs to the first 1-2 s emission of LGRBs. Temporal and spectral studies therefore
point to a common mechanism operating in the first few seconds of any event.
The present work extends this similarity to their late-time emission.

The quasi-linear $w(t_{\rm{pk}})$ evolution shared by long and short GRBs allows us to draw reference lines of equal flare-to-prompt 
energy ratios as a function of $t_{\rm{pk}}/T_{\rm{90}}$ in the $L_{\rm{pk,F}}/\langle L_{\rm{P}}\rangle$ plane 
(blue dotted lines of Fig. \ref{Fig:Lpktpkrenorm}). Flares in the 0.3-10 keV band pass are found to 
emit between $(0.1-100)$\% of the prompt 15-150 keV $E_{\rm{iso,P}}$, with the majority of them lying between the 1\% and 
10\% levels. Little evolution of the flare-to-prompt energy ratio $E_{\rm{iso,F}}/E_{\rm{iso,P}}$ in terms of 
$t_{\rm{pk}}/T_{90}$ can be inferred from the plot (for $t_{\rm{pk}}/T_{90}\lesssim 300$). In particular, SGRB flare
candidates and LGRB flares show comparable flare-to-prompt energy ratios, as reported in Table \ref{Tab:diff}.
This finding provides further support to a physical link between LGRB flares and SGRB flare candidates.

The GRB central engines seem to release comparable fractions of prompt emission energy at late times, 
irrespective of the long or short GRB nature. From Fig. \ref{Fig:Lpktpkrenorm}: 
\begin{equation}
	\frac{L_{\rm{pk,F}}}{\langle L_{\rm{P}}\rangle} = N_1 \Big(\frac{t_{\rm{pk}}}{T_{\rm{90}}}\Big)^{-(1+\alpha)}
	\label{Eq:Lrenorm}
\end{equation}
with $\alpha=-0.1\pm0.1$. The best-fitting rest-frame $w(t_{\rm{pk}})$ relation reads:
\begin{equation}
	\Big( \frac{w}{1+z} \Big)= N_2 \Big(\frac{t_{\rm{pk}}}{1+z}\Big)^{(\beta +1)}
	\label{Eq:w}
\end{equation}
with $\beta=0.2 \pm 0.2$ (Fig. \ref{Fig:wtp}). Equations (\ref{Eq:Lrenorm}) and (\ref{Eq:w}) express common 
properties of LGRB flares and SGRB flare candidates.
The normalisation parameters $N_1$ and $N_2$ 
possibly hide the dependence of $L_{\rm{pk,F}}/\langle L_{\rm{P}} \rangle$ and $w$ on other
parameters. This hidden dependence might be partially responsible for the large scatter of
relation (\ref{Eq:Lrenorm}); $\alpha$ and $\beta$ parametrise the \emph{non}-linear dependence
of $L_{\rm{pk,F}}/\langle L_{\rm{P}} \rangle$ and $w/(1+z)$ on $t_{\rm{pk}}/T_{\rm{90}}$ and $t_{\rm{pk}}/(1+z)$,
respectively. Combining Eq. (\ref{Eq:Lrenorm}) and Eq. (\ref{Eq:w}) it is possible to show that:
\begin{equation}
	\frac{E_{\rm{iso,F}}}{E_{\rm{iso,P}}}= \mathcal{N} \Big(  \frac{t_{\rm{pk}}}{1+z} \Big)^{\beta-\alpha}
	 \Big(\frac{T_{\rm{90}}}{1+z} \Big)^{\alpha}
	 \label{Eq:E}
\end{equation}
where $\mathcal{N}\sim 0.9^2 N_1 N_2$ and $E_{\rm{iso,F}}\sim0.9 L_{\rm{pk}}w/(1+z)$ (valid for a \citealt{Norris05}
flare profile where $w$ is calculated between $1/e$ intensity points 
and with $t_{\rm{r}}=0.5 t_{\rm{d}}$ as found by \citealt{Chincarini10} for LGRB flares).
Equation (\ref{Eq:E}) shows that $E_{\rm{iso,F}}/E_{\rm{iso,P}}\propto [t_{\rm{pk}}/(1+z)]^{\beta-\alpha}$:
a weak dependence of the flare-to-prompt energy ratio on $t_{\rm{pk}}/(1+z)$ 
is expected \emph{if} $\beta-\alpha\sim 0$. From the best-fitting relations, we find that both parameters are
consistent with $0$ at $1\,\sigma$: in particular $\beta-\alpha=0.3\pm0.2$ (consistent with $0$ at
$1.5\,\sigma$ level).  At similar $t_{\rm{pk}}/(1+z)$ a residual dependence on the rest-frame prompt 
duration $T_{\rm{90}}/(1+z)$ is expected to arise from the third term of Eq. (\ref{Eq:E}): this dependence, if
present, would be able to distinguish the population of LGRB flares from SGRB flare candidates detected at the same
rest-frame peak time in terms of $E_{\rm{iso,F}}/E_{\rm{iso,P}}$. However, $\alpha=-0.1\pm 0.1$ (and the relation is highly 
dispersed).
We therefore conclude that the quasi-linear $w(t_{\rm{pk}})$ and $L_{\rm{pk,F}}/\langle L_{\rm{P}} \rangle$ vs. $t_{\rm{pk}}/T_{\rm{90}}$ relations
translate into $E_{\rm{iso,F}}/E_{\rm{iso,P}}$ ratios which, at first order approximation,
show limited dependence on both the properties of the central engine (i.e. duration of the prompt emission) and  
elapsed time from the explosion.

From Fig. \ref{Fig:Ltpk}, $L_{\rm{pk,F}}^{\rm{SGRB}}/L_{\rm{pk,F}}^{\rm{LGRB}}\sim 10^{-2}$ at the same $t_{\rm{pk}}/(1+z)$.
A factor $\sim 100$ is roughly the ratio of the isotropic energy emitted by long and short GRBs during their 
\emph{prompt} $\gamma$-ray emission (Fig. \ref{Fig:Eiso}, inset)\footnote{Note that the same $\sim100$ factor is found as the ratio of 
the isotropic energy emitted by LGRB \emph{flares} and  SGRB \emph{flare} candidates: this is a direct 
consequence of their flare peak luminosity ratio $L_{\rm{pk}}^{\rm{SGRB}}/L_{\rm{pk}}^{\rm{LGRB}}\sim 10^{-2}$, 
coupled to the very similar $w(t_{\rm{pk}})$ evolution.}. LGRB flares and SGRB flare candidates are therefore expected 
to show a similar behaviour in the $L_{\rm{pk,F}}/E_{\rm{iso,P}}$ vs. $t_{\rm{pk}}/(1+z)$ plane (in strict analogy
with the X-ray afterglow scaling found by \citealt{Nysewander09}, their Fig. 6). Equation (\ref{Eq:Lrenorm}) can be easily 
re-arranged into:
\begin{equation}
	\frac{L_{\rm{pk,F}}}{E_{\rm{iso,P}}}=N_1\Big( \frac{t_{\rm{pk}}}{1+z}\Big)^{-(\alpha+1)}\Big( \frac{T_{\rm{90}}}{1+z}\Big)^\alpha.
\end{equation} 
Again, the limited departure of $L_{\rm{pk,F}}/\langle L_{\rm{P}} \rangle$ from a linear relation in $t_{\rm{pk}}/T_{\rm{90}}$ ($\alpha=-0.1\pm 0.1$)
causes LGRB flares and SGRB flare candidates to share the same scaling (at least at first order approximation) .

All the above indications point to a SGRB flare candidates internal origin (Section \ref{Sec:intor})  and establish a 
connection between long and short GRB X-ray variability.
\subsection{The lag-luminosity relation}
\label{SubSec:laglum}

Negligible spectral lag above $\sim25$ keV is the fundamental attribute that makes the prompt $\gamma$-ray 
emission of short bursts different from LGRBs, in addition to their narrower pulses, shorter duration and slightly 
harder emission (\citealt{Norris06} and references therein). The spectral lag is the time difference between
the arrival of high-energy and low-energy photons: in our analysis, a positive value indicates that high energy photons
lead the low energy emission. During the prompt phase of LGRBs the spectral lag $\tau^{\gamma}_{\rm{lag}}$ is 
anti-correlated with the peak luminosity as shown by Fig. \ref{Fig:LagLumFlares} (\citealt{Norris00}; 
\citealt{Ukwatta10}); in contrast, short bursts have small $\tau^{\gamma}_{\rm{lag}}$ (\citealt{Norris06}) and occupy
a separate area of the $L_{\rm{pk}}$ vs. $\tau^{\gamma}_{\rm{lag}}/(1+z)$  parameter space (Fig. \ref{Fig:LagLumFlares} and
\citealt{Gehrels06}). Recently, \cite{Margutti10b} have demonstrated that, in strict analogy to their prompt pulses,
LGRB X-ray flares define a lag-luminosity anti-correlation, where the lag is computed in the X-ray band
(black dots of Fig. \ref{Fig:LagLumFlares}). With the present work we complete the observational picture above,
showing that flare candidates in SGRBs fall off the lag-luminosity relation defined by LGRBs: 
this furthermore supports a robust connection between prompt pulses and flare candidates in short bursts.
At the same time this result points to some differences between LGRB flares and SGRB flare
candidates.

While SGRB prompt pulses are significantly narrower than LGRB pulses (\citealt{Nakar02}; \citealt{Nakar02b}),
flares show instead comparable width in both classes (Fig. \ref{Fig:wtp}). \cite{Hakkila08} showed the 
existence of a lag-width correlation for prompt pulses: the wider the prompt pulse, the longer the lag.
This behaviour has been recently extended to LGRB flares by \cite{Margutti10b}. The similar width
of LGRB flares and SGRB flare candidates implies that the lag-width relation cannot be 
invoked to explain the lag of SGRB flare candidates which are systematically shorter
than expected from the lag-luminosity relation of LGRB flares.

The physical cause of lags in the GRBs prompt emission is not yet understood: variations in the line-of-sight
(\citealt{Salmonson00}); variations of the off-axis angle (\citealt{Ioka01}) and rapid radiation cooling effects
(\citealt{Schaefer04}) are a few of the proposed models. In particular it is at the moment unclear if lags in 
short bursts are small and non-measurable or intrinsically zero. According to the first possibility short and long
GRBs would be powered by a similar, progenitor-independent  physical mechanism, with SGRBs being faster evolving 
versions of LGRBs. The latter would instead point to some intrinsic differences. 
At present it is not possible to observationally distinguish between the two scenarios
\footnote{From the prompt lag-width relation of \cite{Hakkila08}: $\tau^{\gamma,\rm{LGRB}}/w_{\rm{P}}^{\rm{LGRB}}\sim0.01-0.1$.
Using $w_{\rm{P}}^{\rm{SGRB}}\sim0.05$ s as typical value from \cite{Nakar02b},
we have $\tau^{\gamma,\rm{SGRB}}\sim 0.5-5$ ms assuming that the $\tau^{\gamma}/w_{\rm{P}}$ ratio is universal.
We typically resolve lags in SGRBs with a sensitivity of a few ms. This implies that we would be 
barely able to measure lags in the SGRB prompt pulses even if SGRBs were to follow the LGRB $\tau^{\gamma}/w_{\rm{P}}$
relation. }.

In the case of flares, the situation is complicated by the limited and fixed energy band-passes used for the lag
calculation (0.3-1 keV vs. 3-10 keV). The fundamental origin of the observed lag is the spectral evolution 
of a pulse (or flare) profile to lower energies (\citealt{Kocevski03}; \citealt{Margutti10b}). The spectral
peak energy $E_{\rm{peak}}$ decrease in time plays a major role in determining the observed 
spectral evolution and lag value (see \citealt{Margutti10b} for details). As a consequence, if 
the observed $E_{\rm{peak}}$ does not cross the instrumental band pass during the emission, a limited 
spectral evolution will be detected and a short time lag 
determined. The shorter (when compared to what expected from the LGRB flare lag-luminosity relation) 
time lag of SGRB flare candidates 
might be partially a consequence of this observational effect (while possibly being intrinsically larger)
\footnote{Note that the limited brightness of the flare candidates compared to the underlying X-ray continuum
does not allow us to perform a one-to-one comparison with the spectral properties of LGRB flares to quantitatively
check this possibility.}:  
the results from Section \ref{SubSec:flux} and \ref{SubSec:Energetics} suggest that long and short bursts 
basically differ in the intrinsic time scale of central engine evolution (with SGRBs evolving faster). Since for LGRBs
$E_{\rm{peak,F}}^{\rm{LGRB}}/E_{\rm{peak,P}}^{\rm{LGRB}}\lesssim 0.01$ (with 
$E_{\rm{peak,F}}^{\rm{LGRB}}\sim1-3$ keV, observed value, \citealt{Margutti10b}), the faster
evolution of short bursts likely implies  $E_{\rm{peak,F}}^{\rm{SGRB}}/E_{\rm{peak,P}}^{\rm{SGRB}}\ll 0.01$
for flares detected at the same observed $t_{\rm{pk}}$. This result translates into:
$E_{\rm{peak,F}}^{\rm{SGRB}}< 1$ keV considering that 
$E_{\rm{peak,P}}^{\rm{SGRB}}\sim E_{\rm{peak,P}}^{\rm{LGRB}}$ as order of magnitude estimation
(\citealt{Ghirlanda09}).  According to this scenario, $E_{\rm{peak,F}}^{\rm{SGRB}}$ is below the XRT 
band for the majority of the emission, possibly leading to a lag underestimation.
The presence of this observational bias makes the interpretation of the entire lag-luminosity relation
far from being straightforward. We stress that the dependence of the lag-luminosity on the choice of the
 fixed energy bands (both in the rest frame and in the observer frame) should be removed before
 addressing the physical interpretation of the anti-correlation. However, this topic goes beyond the scope
 of this paper and will be addressed in a future work.

With this caveat in mind we note that if the energy $E_{\rm{iso,F}}$  released by flares at 
different $t_{\rm{pk}}$ is similar (as indicated by Section \ref{SubSec:Energetics}), then, considering that  
$L_{\rm{pk,F}}\sim E_{\rm{iso,F}}/w$ with the lag being positively correlated to the $w$ (\citealt{Margutti10b}), 
we would expect $L_{\rm{pk,F}}$ to be anti-correlated with $\tau_{\rm{lag}}^{x}$ as observed for LGRB flares 
of Fig. \ref{Fig:LagLumFlares}: $L_{\rm{pk,F}}=N_{\rm{lag}}^{\rm{F}}(\tau_{\rm{lag}}^{x})^{-1}$. 
In particular the normalisation $N_{\rm{lag}}\propto E_{\rm{iso,F}}$, which implies
$N_{\rm{lag}}^{\rm{LGRB}}\sim\, 100 N_{\rm{lag}}^{\rm{SGRB}}$ (since  
$E_{\rm{iso,F}}^{\rm{LGRB}}\sim 100\, E_{\rm{iso,F}}^{\rm{SGRB}}$, Table \ref{Tab:diff}). This simple argument predicts
the SGRB flare candidates to be off the LGRB flare lag-luminosity relation of a factor $\gtrsim100$
as observed (the $\gtrsim$ inequality accounts for the possible underestimation of the real lag due to the
fixed and limited energy band-passes bias of the previous paragraph).  This finding would support 
the presence of non-mesurable (but still \emph{non-zero}) lags for the short burst prompt emission.
\subsection{The flare candidates internal origin}
\label{Sec:intor}
The above indications point to a link between the properties of flare candidates and prompt pulses in SGRBs
(for LGRBs this was demonstrated by \citealt{Margutti10b}). 
This result would naturally arise if both kind of emission share a common origin: we refer to this possibility as the
\emph{internal origin} interpretation. Theoretical models consistent with this picture can be divided into two categories:
models requiring a late-time GRB central engine activity and models where the central engine
is \emph{not} required to be active at the time of the flare detection.

The second class of models includes the magnetic re-connection interpretation 
(\citealt{Lyutikov03}; \citealt{Giannios06}): flares would originate from residual, late-time magnetic re-connection
events within the original outflow (the \emph{same} ejecta powered the prompt phase) triggered by its deceleration
due to the sweeping-up of the external medium. The deceleration of the original ejecta during the afterglow phase
causes an increase in the size of causally connected regions, thus enabling re-connection of increasingly 
larger magnetic structures. The smooth continuum would be instead afterglow emission
from the shocked external medium.

Alternatively, flares and prompt pulses would automatically share a set of observational properties if they both 
directly arise from the GRB central engine activity (first class of models above).
If this is the case, the central engine would be active on  much longer time-scales than 
previously thought (see e.g. \citealt{Perna06}; \citealt{Rosswog07}; \citealt{Lee09});
at the same time, the  similarity of LGRB flares and SGRB flare candidates
in the  $L_{\rm{pk}}/\langle L\rangle$ vs. $t_{\rm{pk}}/T_{90}$ plane as well as in the
$\Delta F/F$ vs.  $t_{\rm{pk}}/T_{90}$ space  would point to a similar late time evolution
of long and short GRB central engines.

It is not possible to observationally discriminate between the two scenarios using the present set of data.
Careful modelling is required (Margutti et al. in preparation).
\section{Conclusions}
\label{Sec:Con}
This work presents the first comprehensive analysis of flare candidates in SGRBs and provides
a comparison to the observational properties of X-ray flares in LGRBs with the aim to 
draw an observational picture of SGRB X-ray variability \emph{any}
theoretical model is required to explain.

Our analysis shows that the origin of the SGRB X-ray light-curve variability is independent from the
large-scale host galaxy environment and is not correlated to the detected afterglow life-time. 
We find that flare candidates appear in different types of SGRB host galaxy environments
and show no clear correlation with the X-ray afterglow lifetime; flare candidates are detected both in
SGRBs with a bright extended emission (EE) in the soft $\gamma$-rays and in SGRBs
which do not show such component (Sec. \ref{SubSec:EE}).
We cannot exclude that flare candidates originate on top of faint 
(and undetected in the $\gamma$-rays) EEs.
In particular, SGRB flare candidates 
are consistent with being drawn from the LGRB flare population when considering: 
\begin{enumerate}
	\item[1.] The flare to prompt energy ration $E_{\rm{iso,F}}/E_{\rm{iso,P}}$ (Fig. \ref{Fig:Eiso}, Sec. \ref{SubSec:Energetics});
	\item[2.] The relative variability time scale $\Delta t/t <1$  (Fig. \ref{Fig:iokaShort}, Sec. \ref{SubSec:time});
	\item[3.] The rest-frame flare width evolution with time $w(t_{\rm{pk}})$ (Fig. \ref{Fig:wtp}, Sec.  \ref{SubSec:time}).
	\item[4.] The hard-to-soft trend of the emitted radiation  (see e.g. Fig. \ref{Fig:lc100117A}).
\end{enumerate}
The main parameter driving the duration of the episodes of variability is the 
elapsed time from the explosion, with very limited dependence on the different progenitors, environments,
life-times, prompt variability time scales and energy budgets. The origin of the flare $w(t_{\rm{pk}})$ relation
must arise from what is in common for the
long and short burst models. From another perspective this result implies that for $t_{\rm{pk}}>100$ s 
the flare duration is likely to retain no memory of the variability time-scales of the original prompt 
mechanism at work. This would explain why the flare to prompt pulse width ratio is different for long and
short GRBs.

On the contrary, SGRB flare candidates significantly differ from the standard X-ray flare emission 
observed in LGRBs  at the same $t_{\rm{pk}}/(1+z)$ in terms of:
\begin{enumerate}
	\item[5.] Peak luminosity $L_{\rm{pk,F}}^{\rm{SGRB}}\sim 0.01 L_{\rm{pk,F}}^{\rm{LGRB}}$ 
	(Fig. \ref{Fig:Ltpk}, Fig. \ref{Fig:Lpiso}, Sec. \ref{SubSec:Energetics});
	\item[6.] Isotropic energy $E_{\rm{iso,F}}^{\rm{SGRB}}\sim 0.01 E_{\rm{iso,F}}^{\rm{LGRB}}$ 
	(Fig. \ref{Fig:Eiso}, Sec. \ref{SubSec:Energetics});
	\item[7.] Flare to prompt luminosity ratio $L_{\rm{pk,F}}/L_{\rm{pk,P}}$. Flare candidates in SGRBs 
	are $\sim100$ times dimmer than in LGRB;
	\item[8.] Relative variability flux $\Delta F/F$ (Fig: \ref{Fig:iokaShort}): we find $\Delta F/F\sim1$ for all
	SGRB flare candidates (Sec. \ref{SubSec:flux});
	\item[9.] Lag-luminosity relation: like SGRB prompt pulses, flare candidates show shorter lags
	than expected from the lag-luminosity relation of LGRB flares (Fig. \ref{Fig:LagLumFlares}, Sec. \ref{SubSec:laglum}).
\end{enumerate}

However and more importantly, the differences listed at points 5., 6., 7. and 8. above disappear once the
different time scale of evolution of the long and short GRB central engine as well as the different energy scaling 
of the two systems are properly accounted for  (Fig. \ref{Fig:Lpktpkrenorm}, Sec. \ref{SubSubSec:Energetics1} and
\ref{SubSubSec:Energetics2}). This finding provides a connection between the  properties
of the detected SGRB X-ray light-curve variability and LGRB flares, suggesting a common, \emph{internal}
origin. As a result, we conclude that similar dissipation and/or emission mechanisms are likely to be responsible for
the prompt and flare emission in long and short GRBs, with SGRBs being less energetic albeit faster 
evolving versions of the long category.

\section*{Acknowledgments}
We thank the anonymous referee for helpful suggestions that improved the quality of this work.
RM thanks Lorenzo Amati for sharing his data before publication. RM thanks Francesco Massaro,
Mario Guarcello, Paolo Bonfini, Alessandro Paggi, Andrea Marinucci, Vincenzo Cotroneo for their
invaluable help during the writing of the manuscript.
This work is supported by ASI grant SWIFT I/011/07/0, by the Ministry of University and Research of Italy 
(PRIN MIUR 2007TNYZXL), by MAE (Ministry of Exterior), by the University of Milano Bicocca, Italy and
by the ERC advanced research grant "GRBs".

\newpage
\appendix
\section[]{SGRB flares: table}
\begin{table*}
\caption{Best fitting parameters of SGRB flare candidates. From left to right: GRB, redshift; start time ($t_{s}$), amplitude ($A$) and shape parameters
	($\tau_{1},\tau_{2}$) of the best fitting Norris et al. (2005) profile; width evaluated between $1/e$ intensity points; peak time ($t_{\rm{pk}}$); relative variability
	time-scale ($\Delta t/t\equiv w/t_{\rm{pk}}$); relative variability flux $\Delta F/F$: the value $F$ is computed from the best fit of the continuous emission underlying the
	flare candidates (black solid line of  Fig. \ref{Fig:lc100117A}); isotropic equivalent peak luminosity ($L_{\rm{peak}}$) and energy 
	($E_{\rm{iso}}$) computed in the $0.3-10\,\rm{keV}$ observer frame energy band.}
\begin{center}
\resizebox{\textwidth}{!}{\begin{tabular}{llllllllllll}
\hline
GRB & z& $t_s$& $A$& $\tau_1$ & $\tau_2$ & $w$ & $t_{\rm{pk}}$ & $\Delta t/t$ & $\Delta F/F$ &$L_{\rm{pk,F}}^{\rm{SGRB}}$& $E_{\rm{iso,F}}^{\rm{SGRB}}$ \\
        &   & (s)   &($c/s$)& (s)         &  (s)         &  (s)  &   (s)              &                  &                    &($10^{47}\,\rm{erg\,s^{-1}}$)             &  ($10^{48}\,\rm{erg}$)\\ 
\hline
050724 & 0.258 &  $  230.0 $  &  $  12.29 \pm   0.88 $  &  $  12.40 \pm   3.00 $  &  $  35.50 \pm   4.20 $  &  $  65.11 \pm   4.64 $  &  $ 250.98 \pm   1.80 $  &  $   0.26 \pm   0.03 $  &  $   1.72 \pm   0.27 $  &  $   1.49 \pm   0.34 $  &  $   7.16 \pm   1.98 $  \\ 
051210 & --       &  $  107.0 $  &  $   5.70 \pm   1.60 $  &  $ 430.00 \pm 220.00 $  &  $   1.77 \pm   0.87 $  &  $  14.09 \pm   3.52 $  &  $ 134.59 \pm   1.43 $  &  $   0.10 \pm   0.03 $  &  $   1.10 \pm   0.37 $  &  --  &  --  \\
051227 & --       &  $  105.0 $  &  $  12.20 \pm   2.10 $  &  $   8.90 \pm   3.70 $  &  $   5.90 \pm   1.70 $  &  $  14.35 \pm   2.36 $  &  $ 112.25 \pm   0.88 $  &  $   0.13 \pm   0.02 $  &  $   1.62 \pm   0.52 $  &  --  &  --  \\
051227 & --       &  $  122.0 $  &  $   6.00 \pm   0.00 $  &  $  10.00 \pm   0.00 $  &  $  17.40 \pm   2.50 $  &  $  34.94 \pm   4.13 $  &  $ 135.19 \pm   0.96 $  &  $   0.26 \pm   0.03 $  &  $   0.80 \pm   0.21 $  &  --  &  --  \\
060313 & --       &  $  105.0 $  &  $   2.80 \pm   1.20 $  &  $  19.00 \pm  23.00 $  &  $   4.70 \pm   5.30 $  &  $  14.13 \pm   8.87 $  &  $  94.45 \pm   2.21 $  &  $   0.15 \pm   0.09 $  &  $   0.69 \pm   0.32 $  &  --  &  --  \\
060313 & --        &  $  150.0 $  &  $   4.20 \pm   1.70 $  &  $ 111.00 \pm  80.00 $  &  $  10.30 \pm   7.10 $  &  $  38.72 \pm  14.65 $  &  $ 183.81 \pm   4.50 $  &  $   0.21 \pm   0.08 $  &  $   1.03 \pm   0.46 $  &  --  &  --  \\ 
070724A & 0.457 &  $   75.0 $  &  $   7.80 \pm   1.70 $  &  $  25.00 \pm  23.00 $  &  $   7.80 \pm   9.00 $  &  $  22.28 \pm  15.76 $  &  $  88.96 \pm   2.68 $  &  $   0.25 \pm   0.18 $  &  $   0.93 \pm   0.35 $  &  $   3.64 \pm   1.18 $  &  $   5.07 \pm   1.64 $  \\ 
070724A & 0.457 &  $   90.0 $  &  $  12.20 \pm   3.90 $  &  $  42.00 \pm  21.00 $  &  $   5.50 \pm   2.10 $  &  $  19.10 \pm   3.60 $  &  $ 105.20 \pm   1.26 $  &  $   0.18 \pm   0.03 $  &  $   1.46 \pm   0.65 $  &  $   5.68 \pm   2.32 $  &  $   6.72 \pm   2.75 $ \\ 
070724A & 0.457 &  $  150.0 $  &  $   3.79 \pm   0.77 $  &  $  27.00 \pm  17.00 $  &  $  43.00 \pm  18.00 $  &  $  87.80 \pm  21.55 $  &  $ 184.07 \pm   6.34 $  &  $   0.48 \pm   0.12 $  &  $   0.45 \pm   0.17 $  &  $   1.73 \pm   0.74 $  &  $   9.73 \pm   4.16 $  \\
071227 & 0.383 &  $  150.0 $  &  $   4.36 \pm   0.94 $  &  $  80.00 \pm  41.00 $  &  $   7.30 \pm   3.00 $  &  $  27.55 \pm   5.74 $  &  $ 174.17 \pm   2.36 $  &  $   0.16 \pm   0.03 $  &  $   0.80 \pm   0.29 $  &  $   1.13 \pm   0.44 $  &  $   2.03 \pm   0.79 $  \\ 
090607 & --        &  $   89.0 $  &  $   7.60 \pm   1.10 $  &  $  53.00 \pm  31.00 $  &  $  14.00 \pm  12.00 $  &  $  41.49 \pm  22.22 $  &  $ 116.24 \pm   4.84 $  &  $   0.36 \pm   0.19 $  &  $   1.54 \pm   0.45 $  &  --  &  -- \\
090607 & --        &  $  122.0 $  &  $   3.30 \pm   2.90 $  &  $  18.00 \pm  17.00 $  &  $  39.00 \pm  16.00 $  &  $  75.19 \pm  25.52 $  &  $ 148.49 \pm  12.73 $  &  $   0.51 \pm   0.18 $  &  $   0.67 \pm   0.61 $  &  --  &  --  \\ 
100117A & 0.920 &  $  130.0 $  &  $   4.90 \pm   0.73 $  &  $   2.00 \pm   1.80 $  &  $  30.00 \pm   0.00 $  &  $  42.77 \pm   4.98 $  &  $ 137.75 \pm   3.55 $  &  $   0.31 \pm   0.04 $  &  $   0.80 \pm   0.56 $  &  $   9.97 \pm   3.48 $  &  $  21.43 \pm   7.48 $  \\ 
100117A & 0.920 &  $  164.0 $  &  $   7.03 \pm   0.94 $  &  $   6.00 \pm   0.00 $  &  $  42.90 \pm   6.80 $  &  $  67.78 \pm   9.13 $  &  $ 180.04 \pm   1.27 $  &  $   0.38 \pm   0.05 $  &  $   1.14 \pm   0.79 $  &  $  14.09 \pm   5.60 $  &  $  47.45 \pm  18.86 $  \\ 
100117A & 0.920 &  $  200.0 $  &  $   4.30 \pm   1.30 $  &  $ 101.00 \pm  77.00 $  &  $   5.10 \pm   3.50 $  &  $  22.11 \pm   7.86 $  &  $ 222.70 \pm   2.57 $  &  $   0.10 \pm   0.04 $  &  $   0.70 \pm   0.52 $  &  $   8.49 \pm   4.57 $  &  $   8.78 \pm   4.73 $ \\
\hline
\end{tabular}}
\end{center}
\label{Tab:flarepar}
\end{table*}

\section[]{SGRB flares: plots}

\begin{figure*}
\vskip -0.0 true cm
\centering
    \includegraphics[width=0.45\hsize,scale=0.4]{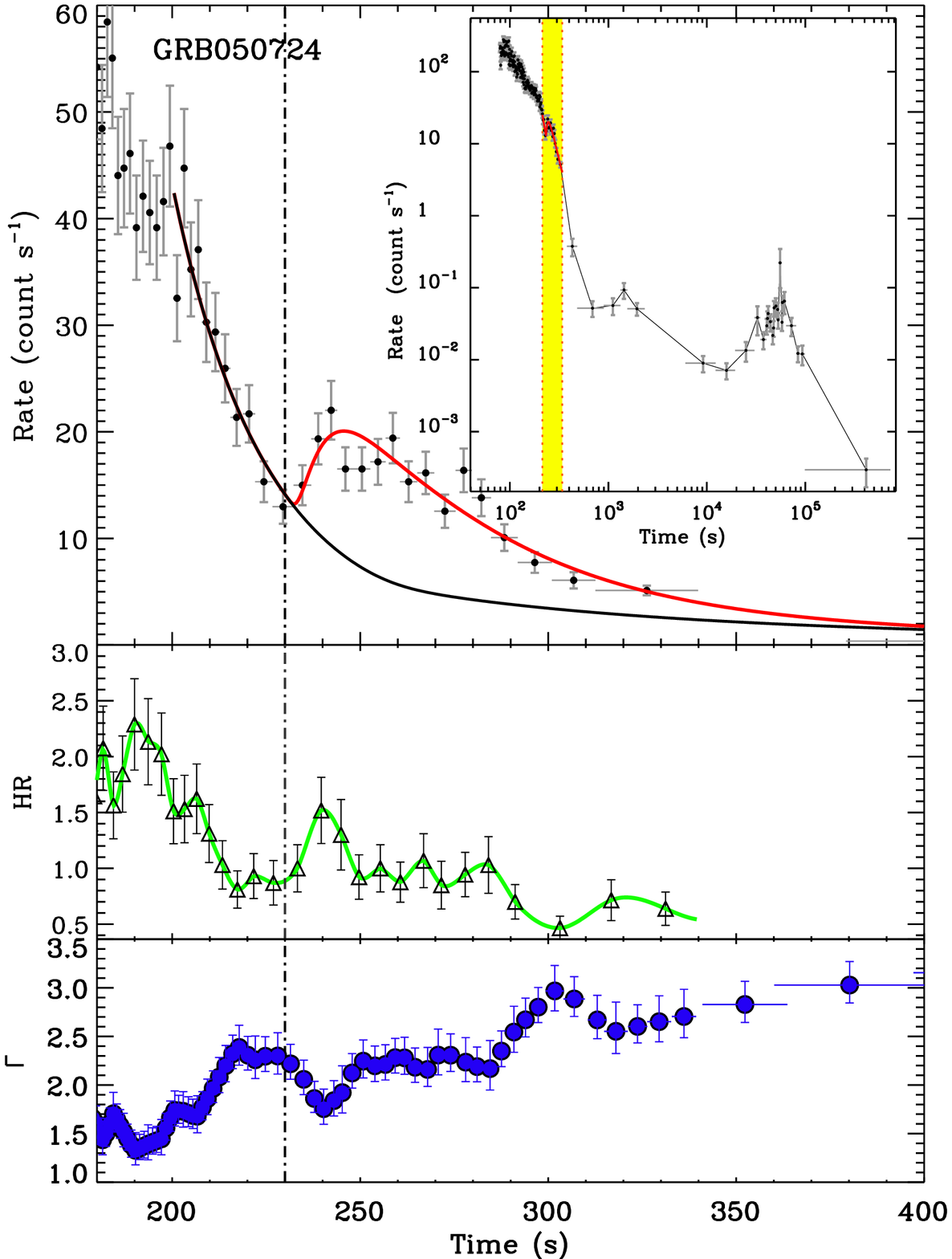}
        \includegraphics[width=0.45\hsize,scale=0.4]{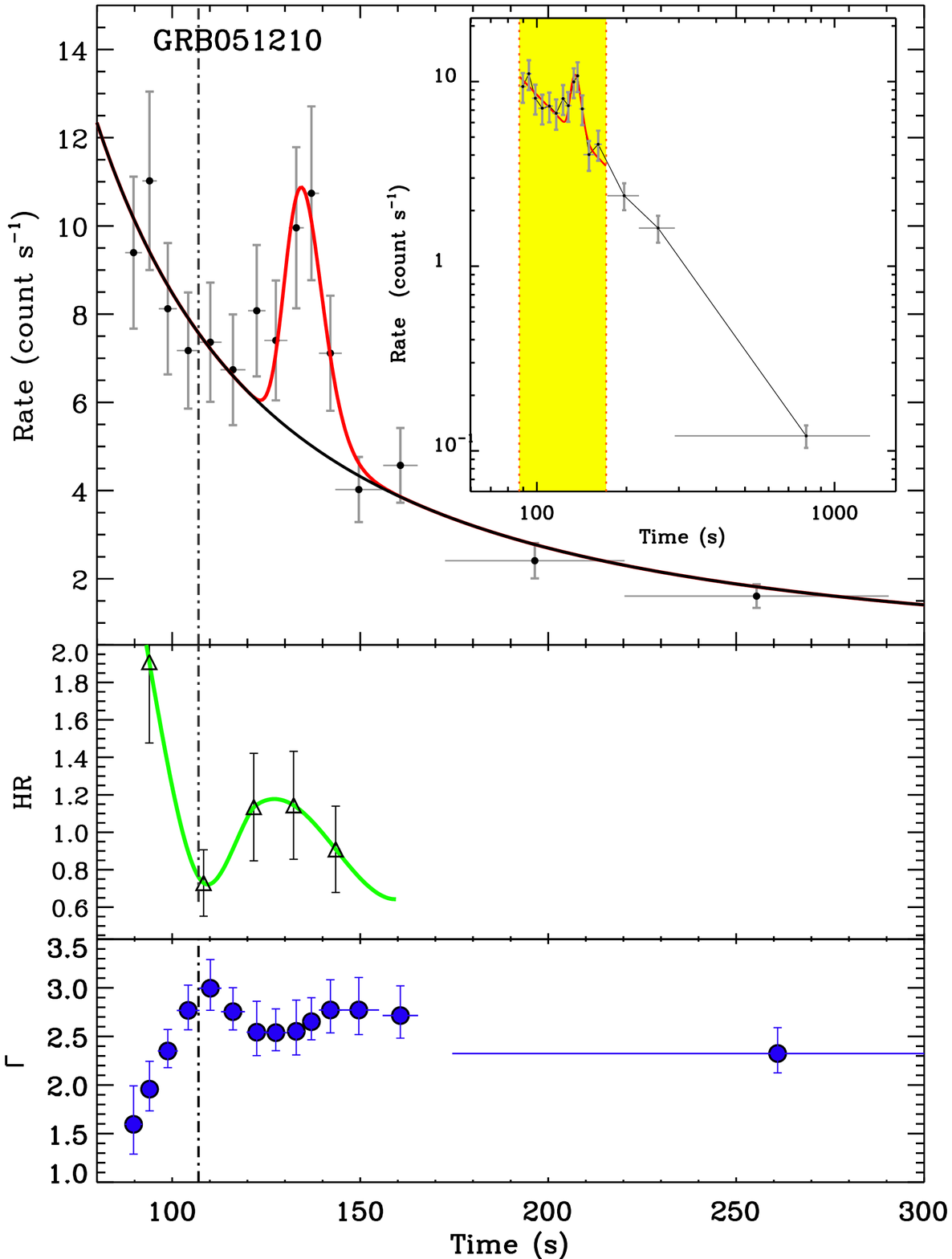}
      \caption{0.3-10 keV count-rate light-curve of GRB\,050724 and GRB\,051210. Black solid line:
       continuous  X-ray emission underlying the flare candidates computed as described in Section \ref{SubSec:XRT};
       dashed lines: best-fitting flare candidate emission; red solid line: best estimate of the total  
      emission. The vertical dot-dashed lines mark the flare candidate onset times.
	 \emph{Inset:} Complete \emph{Swift}-XRT light-curve. The yellow filled area marks the time window for the computation of the CCF lag.
       \emph{Middle panels:} hardness ratio (HR) evolution with time; the HR is computed between 1.5-10 keV (hard band) and 
       0.3-1.5 keV  (soft band). \emph{Lower panels:} Spectral photon index evolution with time as computed by Evans et al. (2010).}
\label{Fig:lc050724051210}
\end{figure*}

\begin{figure*}
\vskip -0.0 true cm
\centering
    \includegraphics[width=0.45\hsize,scale=0.4]{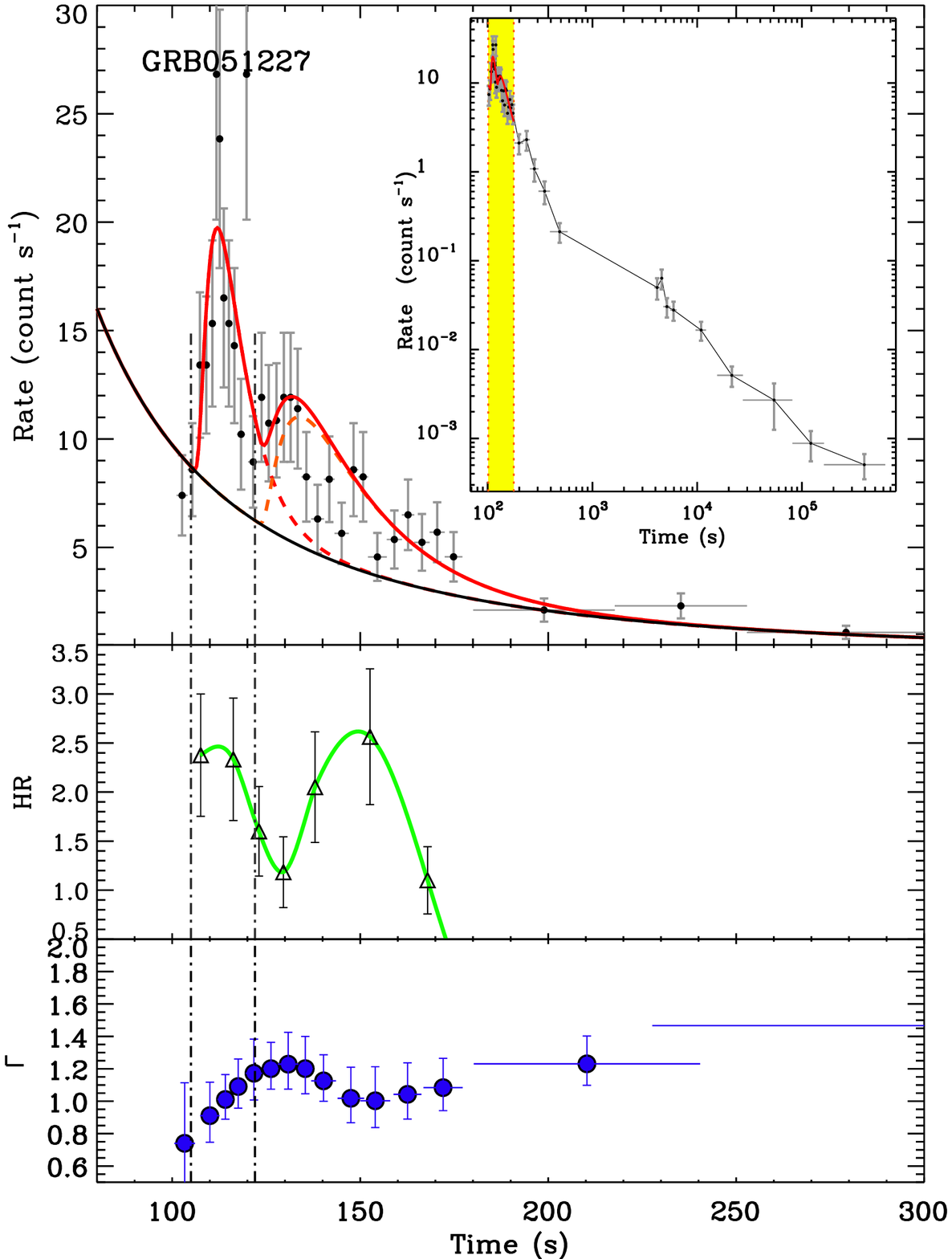}
    \includegraphics[width=0.45\hsize,scale=0.4]{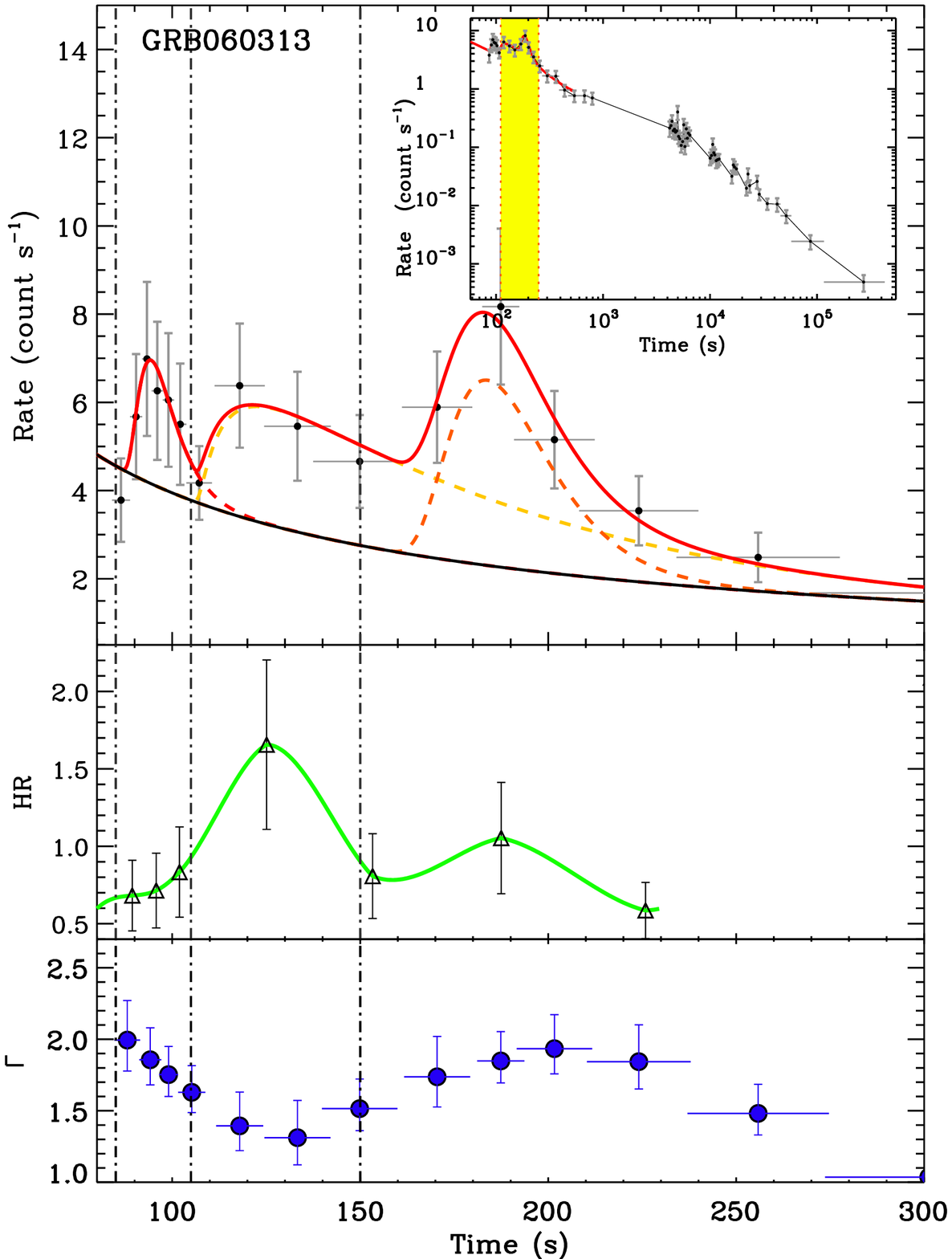}
      \caption{Same as Fig. \ref{Fig:lc050724051210} for GRB\,051227 and GRB\,060313.}
\label{Fig:lc051227060313}
\end{figure*}

\begin{figure*}
\vskip -0.0 true cm
\centering
    \includegraphics[width=0.45\hsize,scale=0.4]{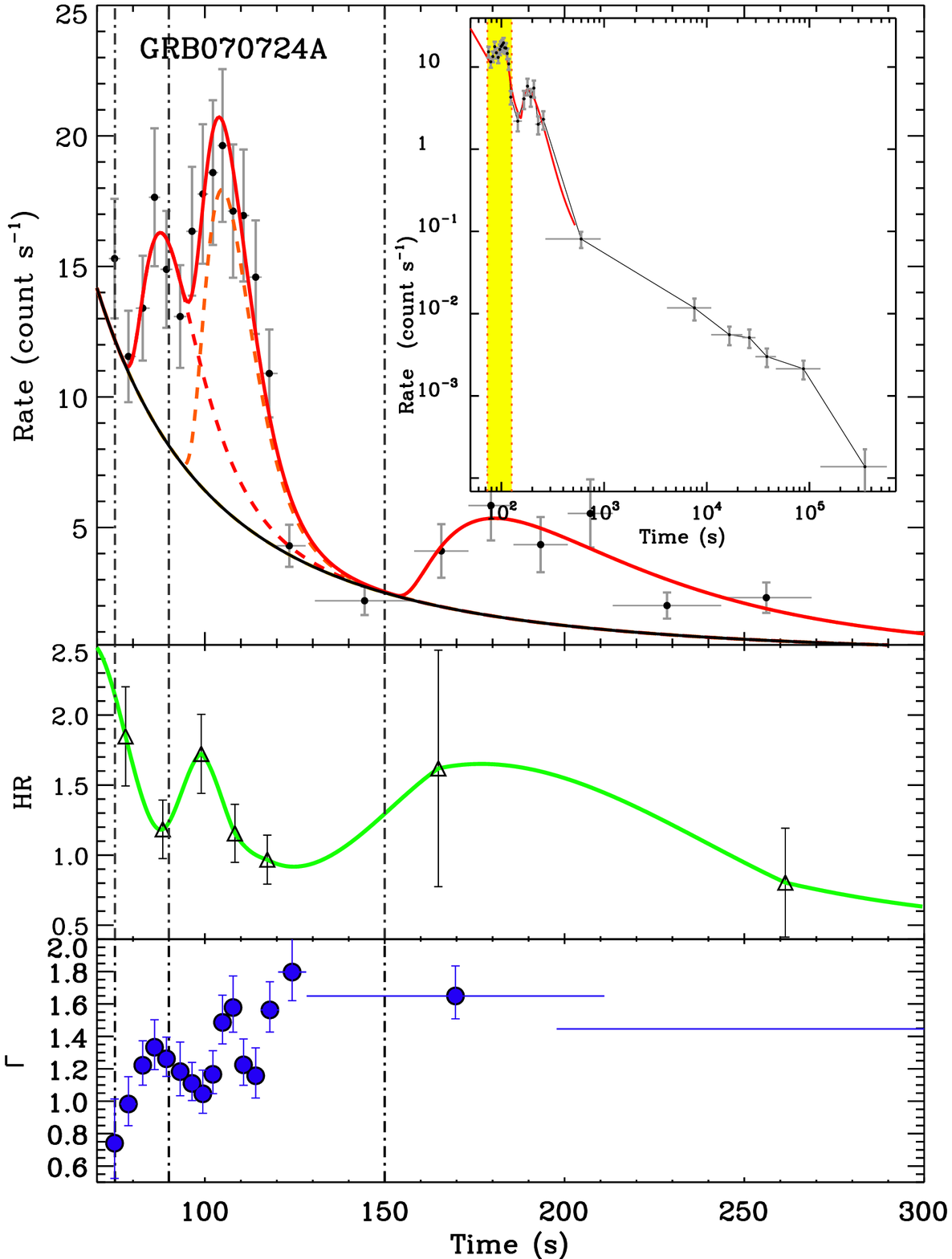}
    \includegraphics[width=0.45\hsize,scale=0.4]{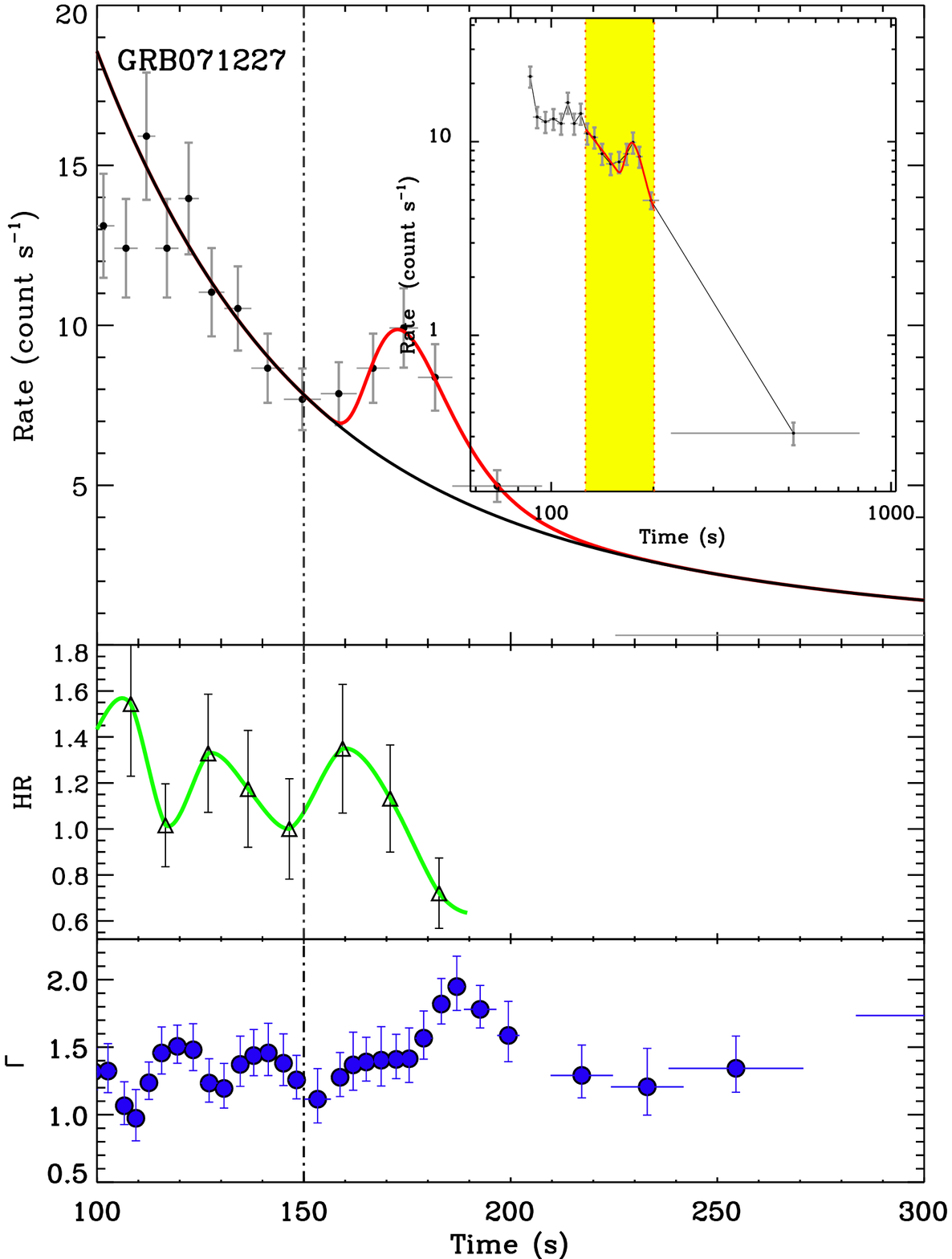}
      \caption{Same as Fig. \ref{Fig:lc050724051210} for GRB\,070724A and GRB\,071227.}
\label{Fig:lclast}
\end{figure*}

\begin{figure}
\vskip -0.0 true cm
\centering
     \includegraphics[scale=0.4]{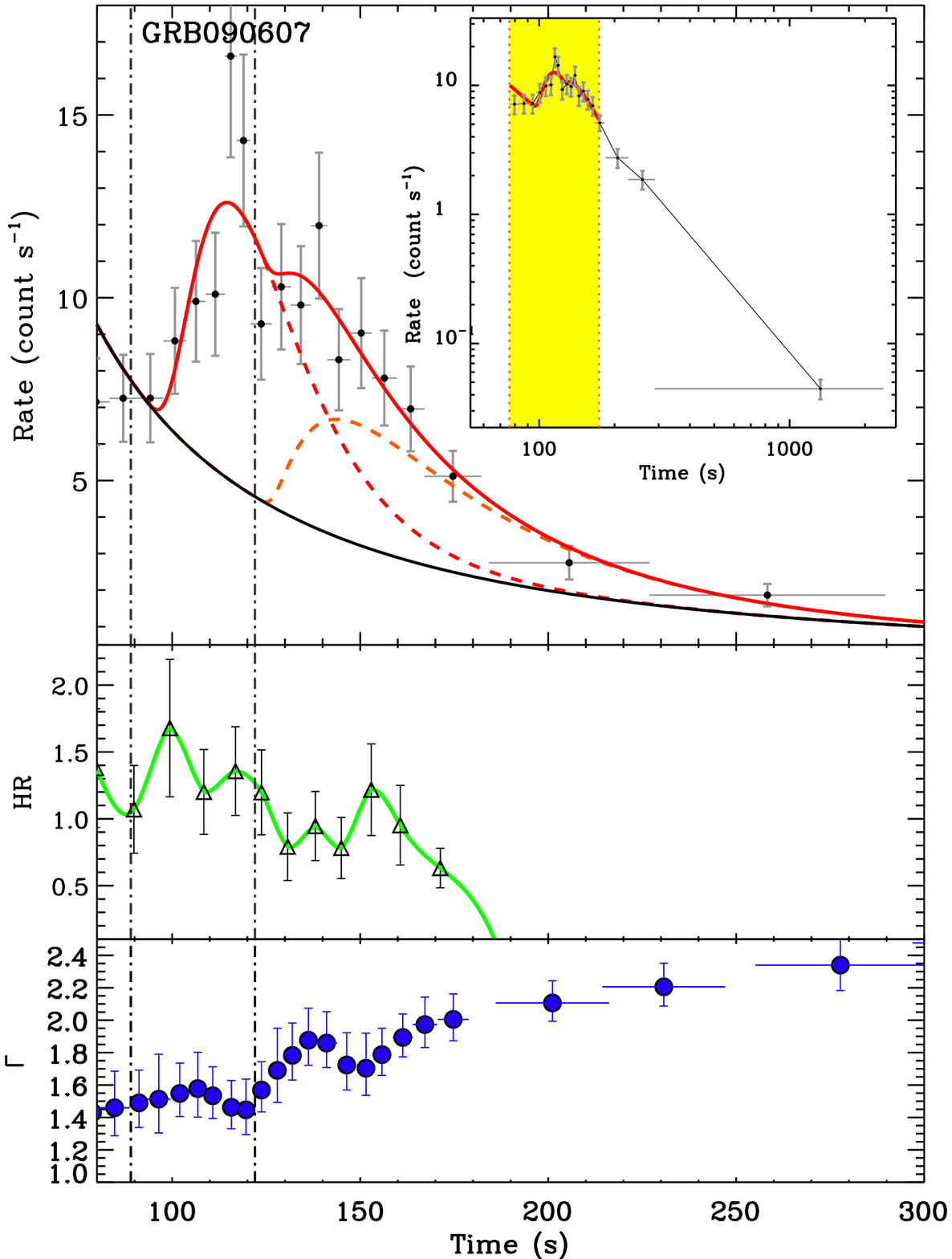}
      \caption{Same as Fig. \ref{Fig:lc050724051210} for GRB\,090607.}
\label{Fig:lclastlast}
\end{figure}

\bsp
\label{lastpage}

\end{document}